\pdfoutput=1
\documentclass{aa} 

\usepackage{natbib}
\usepackage{float}
\usepackage{soul}
\usepackage{amsmath}
\usepackage[draft]{hyperref}
\usepackage{graphicx}
\usepackage{txfonts}
\usepackage{color}
\usepackage{xfrac} 
\usepackage{subfigure}
\usepackage{pdflscape}
\usepackage{lscape}
\usepackage{afterpage}
\usepackage{longtable}

\begin{document}

   \title{Star-forming galaxies at low-redshift in the SHARDS survey}


   \author{A. Lumbreras-Calle
          \inst{1,2}\fnmsep\thanks{\email{alcalle@iac.es}}
          \and
          C. Muñoz-Tuñón\inst{1,2}
          \and
          J. Méndez-Abreu\inst{1,2,3}
          \and
          J. M. Mas-Hesse\inst{4}
          \and
          P.G. Pérez-González\inst{5}
          \and
          B. Alcalde Pampliega\inst{5}
          \and
          P. Arrabal Haro\inst{1,2}
          \and
          A. Cava\inst{6}
          \and
          H. Domínguez Sánchez\inst{7,8}
          \and
          M. C. Eliche-Moral\inst{5}
          \and
          A. Alonso-Herrero\inst{9}
          \and
          A. Borlaff\inst{1,2}
          \and
          J. Gallego\inst{5}
          \and
          A. Hernán-Caballero\inst{5}
          \and
           A. M. Koekemoer\inst{10}
          \and
         L. Rodríguez-Muñoz\inst{11}}

   \institute{Instituto de Astrofísica de Canarias, Calle Vía Láctea s/n, E-38200 La Laguna, Tenerife, Spain
         \and
             Departamento de Astrofísica, Universidad de La Laguna, E-38205 La Laguna, Tenerife, Spain
         \and 
             School of Physics and Astronomy, University of St Andrews, SUPA, North Haugh, KY16 9SS, St Andrews, UK
         \and
             Centro de Astrobiología, CSIC / INTA, Ctra. de Torrejón a Ajalvir, 4 km, 28850 Torrejón de
Ardoz, Madrid, Spain
\and 
    Departamento de Astrofísica y Ciencias de la Atmósfera, Facultad de CC. Físicas, Universidad Complutense de Madrid, E-28040, Madrid, Spain
\and     
    Observatoire de Genève, Université de Genève, 51 Ch. des Maillettes, 1290 Versoix, Switzerland
\and
GEPI, Observatoire de Paris, CNRS, Univ. Paris Diderot; Place Jules Janssen, 92190 Meudon, France
\and
Department of Physics and Astronomy, University of Pennsylvania, Philadelphia, PA 19104, USA
\and
Centro de Astrobiología (CAB, CSIC-INTA), ESAC Campus, E-28692 Villanueva de la Cañada, Madrid, Spain
\and
Space Telescope Science Institute, 3700 San Martin Drive, Baltimore, MD 21218, USA 
\and 
   Dipartimento di Fisica e Astronomia, Università di Padova, vicolo dell'Osservatorio 2, I35122 Padova, Italy}
   \date{Received ; accepted }

 \abstract
 {The physical processes driving the evolution of star formation (SF) in galaxies over cosmic time still present many open questions. Recent galaxy surveys allow now to study these processes in great detail at intermediate redshift ($0 \leq z \leq 0.5$).}
{ We build a complete sample of star-forming galaxies and analyze their properties, reaching systems with low stellar masses and low star formation rates (SFRs) at intermediate-to-low redshift.}
{We use data from the SHARDS multiband survey in the GOODS-North field. Its depth  (up to magnitude $\langle m_{3\sigma}\rangle\sim26.5$) and its spectro-photometric resolution ($R\sim50$) provides us with an ideal dataset to search for emission line galaxies (ELGs). We develop a new algorithm to identify low-redshift ($z$<0.36) ELGs by detecting the [OIII]5007 and H$\alpha$ emission lines simultaneously. We fit the spectral energy distribution (SED) of the selected sample, using a model with two single stellar populations.}
{We find 160 star-forming galaxies for which we derive equivalent widths (EWs) and absolute fluxes of both emission lines. We detect EWs as low as 12 \text{\AA}, with median values for the sample of $\sim$ 35 \text{\AA} in [OIII]5007 and $\sim$ 56 \text{\AA} in H$\alpha$, respectively. Results from the SED fitting show a young stellar population with low median metallicity (36\% of the solar value) and extinction ($A_V \sim$ 0.37), with median galaxy stellar mass $\sim$ 10$^{8.5}$ M$_{\odot}$. Gas-phase metallicities measured from available spectra are also low. ELGs in our sample present bluer colours in the UVJ plane than the median colour-selected star-forming galaxy in SHARDS. We suggest a new (V-J) colour criterion to separate ELGs from non-ELGs in blue galaxy samples. In addition, several galaxies present high densities of O-type stars, possibly producing galactic superwinds, which makes them interesting targets for follow-up spectroscopy.}{
We have demonstrated the efficiency of SHARDS in detecting low-mass ELGs ($\sim$2 magnitudes deeper than previous spectroscopic surveys in the same field). The selected sample accounts for 20\% of the global galaxy population at this redshift and luminosity, and is characterized by young SF bursts with sub-solar metallicities and low extinction. However, robust fits to the full SEDs can only be obtained including an old stellar population, suggesting the young component is built up by a recent burst of SF in an otherwise old galaxy. }

   \keywords{Galaxies: star formation -- Galaxies: photometry  -- Galaxies: fundamental parameters -- Galaxies: stellar content -- Galaxies: starburst
               }

   \maketitle

\section{Introduction}
Understanding the key physical processes that govern the formation and evolution of galaxies is one of the most active and debated topics in modern astrophysics. Star formation (SF) is one of them, and its evolution along the history of the Universe still presents several open questions. 

A plethora of studies have approached this problem by tracing the evolution with redshift of the star formation rate density, obtained as the average star formation rate (SFR) per unit comoving volume (see \citealt{2014ARA&A..52..415M} for a review). These works provide a remarkably consistent picture of the cosmic star formation history (SFH) with an initial rising trend that peaks at $z\sim 2, $ followed by a decline of an order of magnitude down to the values measured locally. We lack, however, a full understanding of the underlying processes that shape this behavior. From simulations, we know that, at least at $z<$2, the cosmic SFH is governed by the physical characteristics of the SF in galaxies (gas fraction, feedback, efficiency, etc.; see \citealt{2010MNRAS.402.1536S}).

The revolutionary imaging performed by the Hubble Space Telescope (HST) on the Hubble Deep Field \citep{1996AJ....112.1335W} enabled the first high-resolution studies of morphology and other properties of galaxies across cosmological times \citep{1996MNRAS.279L..47A,1996AJ....112..359V,2004ApJ...604L..21E,2004ApJ...603...74E}. Interestingly, they revealed a different panorama of  SF processes at low and high redshifts. In particular, distant star-forming galaxies are dominated by clumpy morphologies, with SF occurring in kiloparsec-size regions. These regions are orders of magnitude larger and more massive (M$_{\star}\sim10^7-10^9$, or larger) than typical local HII regions, although similar to those in local luminous and ultraluminous infrared galaxies \citep[see e.g.,][]{2006ApJ...650..835A,2012A&A...541A..20A,2016A&A...590A..67P}. However, in both simulations \citep[e.g.,][]{2015MNRAS.453.2490T,2016ApJ...819L...2B} and observational studies \citep{2017ApJ...836L..22D}, the intrinsic properties of these distant star-forming clumps, especially their mass and size upper limits, have been revisited, suggesting smaller typical values.

The formation mechanism of these massive high-redshift clumps is yet unclear, with several options being proposed: disk fragmentation in gravitationally unstable disks \citep{1999ApJ...514...77N,2007ApJ...670..237B,2008ApJ...688...67E}, intense inflow of cool gas, able to provide the high gas surface densities leading to the disk instabilities \citep{2009ApJ...703..785D,2010Natur.467..811C,2013ApJ...767...74S,2014A&ARv..22...71S}, or ex-situ clumps accreted by minor mergers into the galaxy disk \citep{2014MNRAS.443.3675M}. The different theoretical explanations can be tested studying the properties of both the star-forming clumps and the host galaxy. 

Previous studies by our group \citep{2016A&A...592A.122H} have compiled and analyzed starburst galaxies at $z$<0.5 in the COSMOS survey. Their results show that starburst galaxies are also clumpy at that redshift, with SF knots showing properties somewhat intermediate between those of high-redshift and local starbursts, and with the more massive knots located closer to the galaxy center. These results support the predictions of numerical simulations, claiming that clumps are caused by violent disk instabilities, that may coalesce together and form the central giant clumps/bulges \citep{1999ApJ...514...77N,2007ApJ...670..237B,2008ApJ...688...67E}. 

The main objective of this paper is to extend the analysis of emission line galaxies to lower masses and lower SFRs beyond the local Universe. We explore the characteristics of such a population, and how they relate with their higher-mass, higher-SFR counterparts. We use data from the Survey for High-z Absorption Red and Dead Sources (SHARDS) \citep{2013ApJ...762...46P}, a deep multi-band photometric survey with continuous optical spectral coverage and medium band filters in the GOODS-North field. Its depth and the narrowness of its filters allows us to measure low equivalent width (EW) lines and low-mass emission line galaxies (ELG). The sample detection is based on the simultaneous identification of galaxies with H$\alpha$ and [OIII]5007 emission lines. We then analyze the main integrated properties of their stellar populations via spectral energy distribution (SED) fitting.

We have focused our work mostly on the SHARDS data for consistency, to minimize problems of aperture matching, different filters sizes, and absolute calibrations with respect to other surveys covering the same area. We have included only data from the ALHAMBRA (Advanced Large Homogeneous Area Medium Band Redshift Astronomical) survey and the GALEX (Galaxy Evolution Explorer) mission to complement our SHARDS photometry to better derive the properties of the young population in these galaxies, since the stellar continuum below 4000 {\AA} is important to constrain them.

This work is structured as follows: In Section \ref{data} we present the observational databases used in this paper, in Section \ref{detection_lines} we describe the procedure to detect emission lines and in Section \ref{sample} we describe the resulting sample. In Section \ref{SEDfit} we outline the procedure followed to perform the SED fitting analysis, in Section \ref{spectra} we describe the analysis of the available spectra and in Section \ref{sedfit_results} we show and present the results of both analyses. Finally in Section \ref{conclusions} we summarize our conclusions. 

Throughout this paper we consider standard $\Lambda$CDM cosmology, with $\Omega_{\Lambda}$=0.7, $\Omega_{M}$=0.3 and $H_{0}=70$ km s$^{-1}$ Mpc$^{-1}$. Every mention of EW (unless specified) refers to rest-frame EW. All references to magnitudes correspond to AB magnitudes.

\section{Observational databases}
\label{data}
The main source of observational data for this study is the SHARDS survey (Pérez-González et al. 2013, Barro et al. in preparation). SHARDS was performed using the OSIRIS (Optical System for Imaging and low-Intermediate-Resolution Integrated Spectroscopy) instrument, at the 10.4 m Gran Telescopio de Canarias (GTC) at the Observatorio del Roque de los Muchachos, in La Palma. It consists of very deep imaging of the Great Observatories Origins Deep Survey - North field (GOODS-N, \citealt{2004ApJ...600L..93G}), reaching down to a limiting magnitude $m_{lim} \sim 26.5$ at $3\sigma$  (see \citealt{2013ApJ...762...46P} for details). It made use of 25 contiguous medium band filters, with full width half maximum (FWHM) of $\sim 170 \text{\AA}$, reaching equivalent spectroscopic resolution of $R\sim 50$ over the whole field of $\sim$ 130 arcmin². The wavelength range covers the range 5000 \text{\AA} to 9500 \text{\AA}. The SHARDS observations were taken during a period of $\sim$ 200 hours of dark time with seeing better than 1". 

  \begin{figure*}

   \centering
   \subfigure{\includegraphics[width=0.81\textwidth,keepaspectratio]{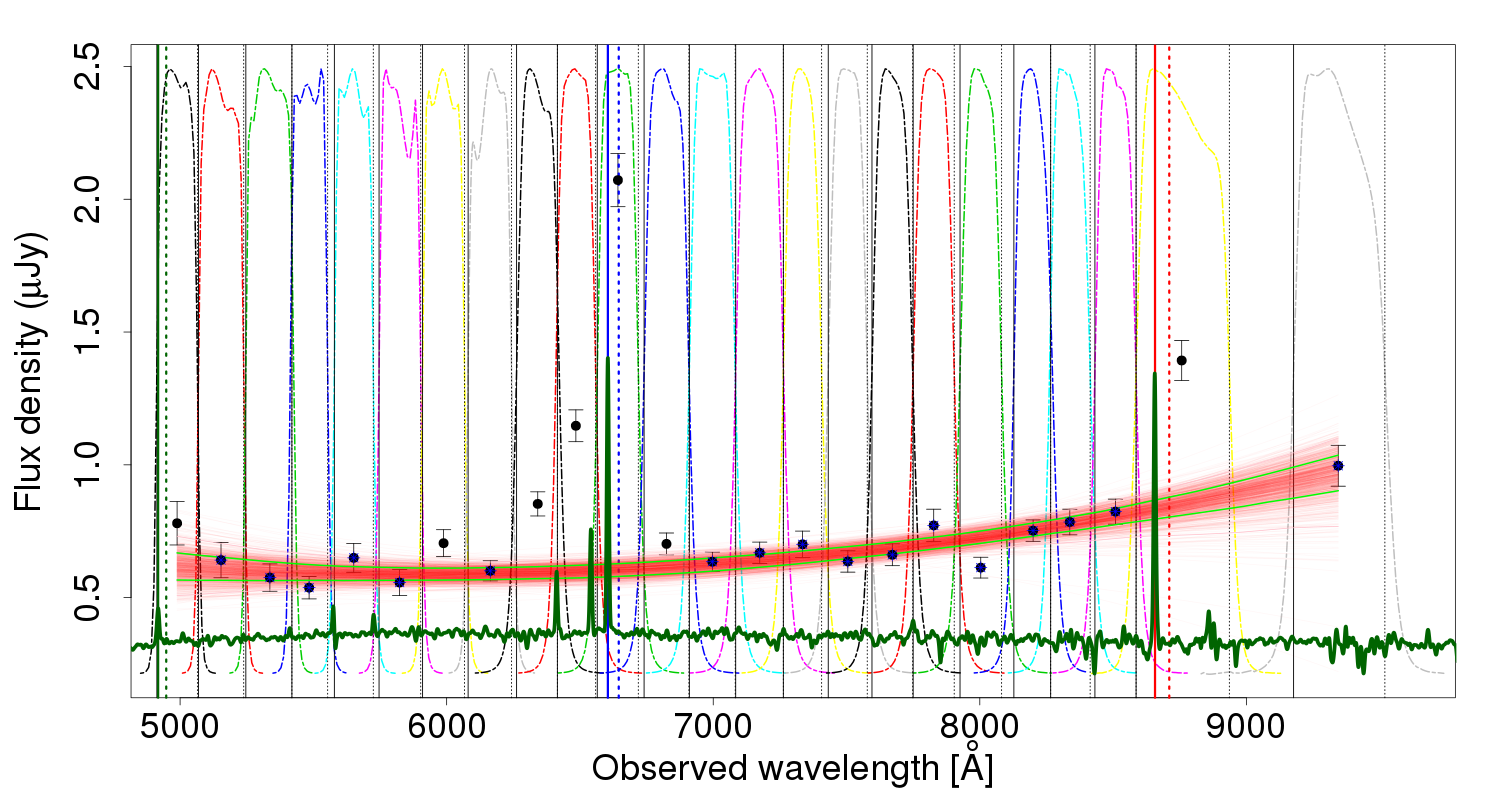}}
\subfigure{\includegraphics[width=0.4\textwidth,keepaspectratio]{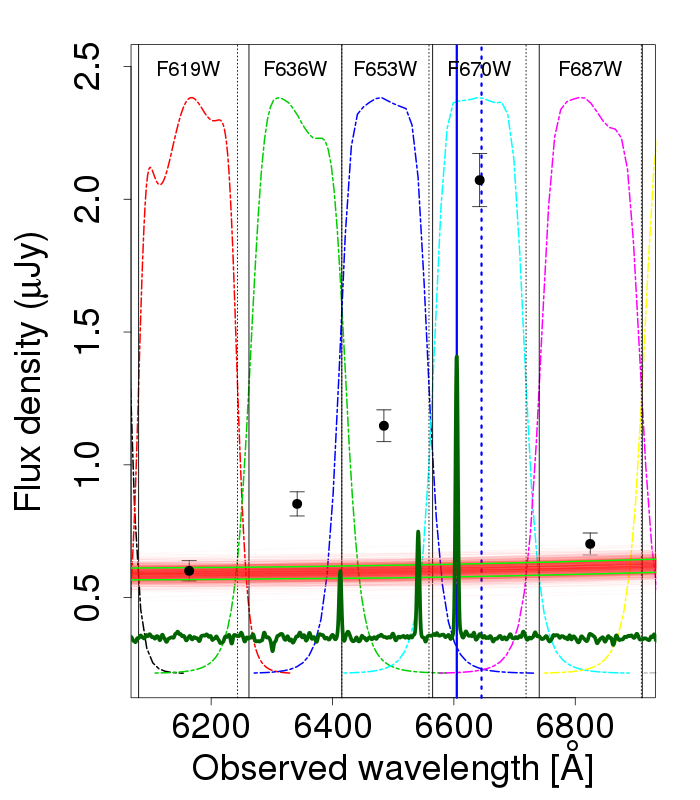}}
\subfigure{\includegraphics[width=0.4\textwidth,keepaspectratio]{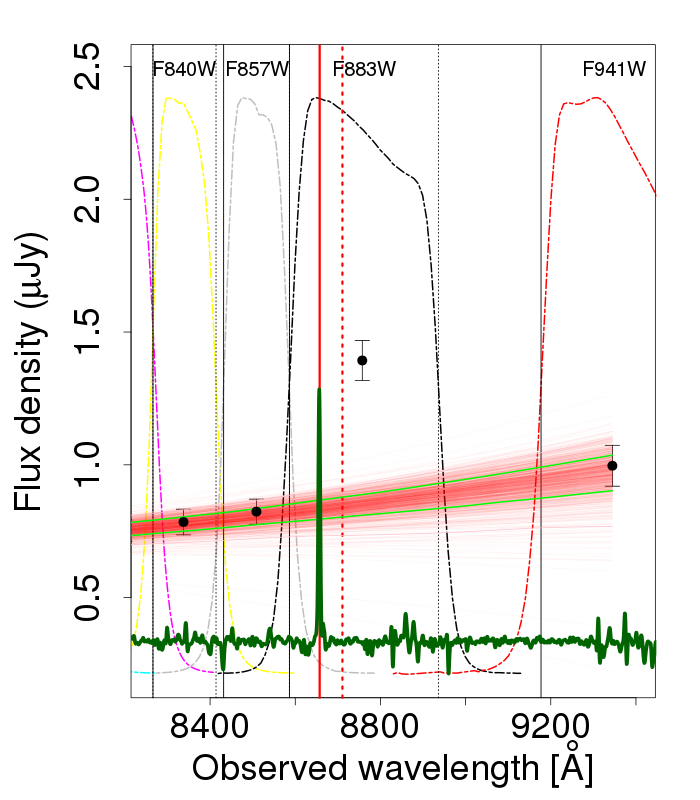}}

      \caption{\textit{Top:} SED of an ELG in our sample (SHARDS10003416) in black dots, overplotted on the transmission profile of SHARDS filters. The longslit spectrum (from the TKRS survey) is also shown in dark green. The 1000 bootstrap continuum fits are shown in red, with the $\pm1\sigma$ limits in green. Vertical dashed and continuous black lines indicate the lowest and highest wavelengths, respectively, for which each filter reaches 50\% transmission rate. Vertical red and blue lines represent the expected wavelength of the H$\alpha$ and [OIII]5007 lines, respectively, considering spectroscopic redshift (continuous lines) and photometric redshift derived in this work (dashed lines). Notorious flux excess can easily be seen in several filters, corresponding to emission lines. 
     \textit{Bottom left:} Zoom into the [OIII]5007,4959 emission lines, using the same colors and symbols as the upper panel.
     \textit{Bottom right:} Zoom into the H$\alpha$ line spectral region.}
         \label{fig:SED}
   \end{figure*}

In order to identify star-forming galaxies, we used in this work the photometric catalog created using data from the 24 filters observed and reduced as of January 2016. The catalog gives the photometric measurements and uncertainties in each of the filters, for the best elliptical aperture, as determined by SExtractor \citep{1996A&AS..117..393B}. In addition to the photometric uncertainty, we considered that of the absolute calibration of each filter \citep{2013ApJ...762...46P}, adding them in quadrature.

Since SHARDS was designed to target galaxies at higher redshift than those in the present study, its wavelength range falls short of covering the Balmer break region ($\sim$ 4000 \text{\AA} rest frame). This spectral range provides fundamental information about the stellar populations of the galaxies. To extend the wavelength range, we also use data from the ALHAMBRA survey \citep{2014MNRAS.441.2891M}. ALHAMBRA is a spectro-photometric survey using 20 contiguous medium band (FWHM $\sim$ 300\AA) filters in the wavelength range 3500\AA<$\lambda$<9700\AA. We considered all galaxies in SHARDS as our parent sample, but ALHAMBRA only covers 45\% of the GOODS-N field.

To further constrain the ultra-violet range, we used data from the GALEX space telescope \citep{2014yCat.2335....0B}. Both filters, FUV ($\lambda_{mean}= 1528${\AA}) and NUV($\lambda_{mean}= 2371 ${\AA}) were used, with a limiting magnitude in both bands of $m_{lim}\sim$ 25. Data in this wavelength range help us to constrain the young stellar population properties, in particular its extinction.

  Spectroscopic data is also available for a subset of galaxies in our sample in the public releases of the TKRS \citep[Team Keck Redshift Survey,][]{2004AJ....127.3121W} and DEEP3 \citep[Deep Extragalactic Evolutionary Probe 3,][]{2011ApJS..193...14C} surveys. They were performed from the Keck Telescopes in Hawaii, using a 600 $mm^{-1}$ grating at the DEep Imaging Multi-Object Spectrograph (DEIMOS). We also used the spectroscopic redshifts database in \cite{2008ApJ...689..687B}, as well as one-dimensional spectra from the same study, kindly provided by S. Barger.

The GOODS-N cosmological field has also been observed by multiple HST surveys, with broadband imaging and infrared (IR) grism spectroscopy. These include CANDELS \citep{2011ApJS..197...35G,2011ApJS..197...36K} and the 3D-HST survey \citep{2012ApJS..200...13B, 2014ApJS..214...24S}, with photometric redshifts determined from IR grism and photometric observations. We made use of these data as an independent redshift value for our targets, as well as a more precise measurement of the size of the galaxies than what is possible from ground-based observations.

\section{Detection of emission line galaxies}
\label{detection_lines}

One of the preferred methods for the detection of star-forming galaxies is the identification of nebular emission lines, associated with HII regions surrounding young stellar clusters (and therefore, tracers of recent SF).
The SHARDS dataset provides an excellent benchmark to analyze the SED of galaxies thanks to its large wavelength coverage, good equivalent spectral resolution, high depth, and very good image quality. This allows us to accurately reconstruct the SED of individual galaxies, in particular to detect the presence of emission lines as a flux excess in a given filter (thanks mainly to the narrowness of the filters). In addition, it is less time consuming and reaches fainter and more numerous targets than spectroscopic surveys, and does not need a pre-selected sample.
In the present work, we simultaneously searched for both H$\alpha$ and [OIII]5007 emission lines in each galaxy. This was done in order to avoid spurious detections and to improve the precision and robustness of the photometric redshift determinations (see \citealt{2016A&A...592A.122H} for a similar approach with a different dataset, and \citealt{2015ApJ...812..155C} for a single-line approach in the same dataset).
Before running our detection algorithm, we made a redshift cut to the main SHARDS sample, taking only galaxies with photometric redshift (in both SHARDS and 3D-HST catalogs) lower than 0.36. This was necessary to ensure that the emission lines we detect are H$\alpha$ and [OIII], and not a different pair of lines with a similar wavelength distance between them, in particular [OIII]5007-[OII]3727. The limit at $z  < $ 0.36 is a result of the wavelength limit of F883W35, the reddest filter used in the detection of H$\alpha$. We did not consider detections in the bluest or reddest filters available (F500W17 and F941W33) to ensure that the continuum estimation under the lines was not an extrapolation, which would imply high uncertainty. This left us with a parent sample of 1823 galaxies.

The measured emission in SHARDS filters can be contaminated by other emission lines, most notably [NII]6583 and [NII]6549 for H$\alpha$ and [OIII]4959 and H$\beta$ for [OIII]5007. Only when H$\alpha$ emission is detected in the reddest filter considered, F883W35, do we consider that [SII]6718+6732 contamination may also be present. This issue is addressed in Section \ref{EWcalc}. 

In the following subsections, we review in detail the process to detect the emission lines.

\subsection{ELG detection procedure}
\label{em_lin_detec}

For each galaxy, the SHARDS SED is analyzed to detect H$\alpha$ and [OIII]5007 emission lines, deriving a new estimate of the redshift and computing their EWs and fluxes. The procedure runs as follows.

In order to determine if a given filter shows an excess of flux it is necessary to define a baseline, the stellar continuum. To estimate it, we first performed a sigma-clipped second order polynomial fitting procedure to remove potential emission lines (and poor photometric measurements) in SHARDS photometric points. To properly assess the uncertainty in the continuum, 1000 Bootstrap and Monte Carlo simulations were performed over the remaining points after the sigma-clipping, fitting again a quadratic function to the points in each simulation. The resulting parameters provide an empirical distribution of probability for the value of the continuum at each wavelength, resulting in a robust estimation of its uncertainty. A series of tests using higher-order polynomials or splines were performed to check the accuracy of our quadratic assumption, but they did not show significant improvement in the quality of the fit, so we kept the quadratic fit for simplicity. Figure \ref{fig:SED} shows the SED of one galaxy in our sample, with the transmission profile of SHARDS filters overplotted, as well as the continuum fits.

The detection of emission lines over the continuum is performed as follows: First, the difference in flux between each filter and the continuum is computed. Then, after masking the two filters with the highest values (assumed to be the [OIII]5007 and H$\alpha$ emission lines) we derive the root mean square error ($RMSE$) of these differences around the continuum fit. This value is used to define the noise of the SED. We assume that the filter with the highest emission excess over the continuum corresponds to H$\alpha$ (and thus derive a tentative redshift). It must exceed a certain threshold ($1.5 \times RMSE$) and have a central wavelength longer than 6450 \text{\AA}. Then, the code searches for an excess in flux in the filter where the [OIII]5007 line should lie at the tentative redshift of the galaxy. After this (even if it succeeded) it looks for the other possible case (that the highest emitting filter is the [OIII]5007 line), and then searches for H$\alpha$. This process is repeated for the second brightest filter. The pair of filters that shows the highest excess over the $RMSE$ is considered to be the correct [OIII]-H$\alpha$ match. 

\subsection{Robustness of the detection}
\label{sec_robust}
After the continuum and line detection procedures are performed, another step is necessary to calculate a more precise parameter of signal-to-noise ratio (S/N) for the measured lines, and then preserve only the statistically significant cases.

This is archived analyzing the output of the continuum simulations described in Section \ref{em_lin_detec}. For each filter, we compute the width of the distribution, $\Delta C_i$, as the range that holds the central 68.27\% of the continuum values ($\pm 1\sigma$) at the central wavelength of filter $i$. We also define the upper limit of the continuum $C_{U84,i}$ ($C_{+\sigma}$) as the value that leaves below 84.135\% of the continuum simulations for that filter. We also take $F_i$ and $errF_i$ from the SHARDS catalog (the value of the flux in that filter and its error, respectively). The value we need to consider in both cases is the difference between the filter emission and the upper limit of the continuum, $F_i-C_{U84,i}$, not just $F_i$, since we are not interested in the absolute S/N of the flux, but in its S/N \textit{above the continuum}. Taking all this into account, we derive two parameters related to the S/N: 
\begin{itemize}
\item $P_{cont}$: The difference between the flux in a particular filter and the $C_{U84,i}$ value for that filter divided by the width ($\Delta C_i$) of the continuum distribution. Placing a threshold in this parameter ensures that the suspected emission line is not an artifact caused by a noisy continuum.
\begin{equation}
P_{cont}=\frac{F_i-C_{U84,i}}{\Delta C_i}
.\end{equation}
\item $P_{phot}$: The difference between the flux in a particular filter and the $C_{U84,i}$ value for that filter, divided by the error of the flux in that filter ($errF_i$). Placing a threshold in this parameter ensures that the suspected emission line is not an artifact caused by a noisy photometric point.
\begin{equation}
P_{err}=\frac{F_i-C_{U84,i}}{errF_i}
.\end{equation}
\end{itemize}

A threshold in both parameters is necessary to ensure that the filter shows significant emission. We gathered samples using different thresholds (1, 1.5, and 2), and decided to keep the threshold as 1.5 for both parameters. Visual inspection of the limiting cases of the three samples showed that the 1.5 threshold rejected very little clear cases while keeping a good confidence in the significance of the accepted emitters. We note that this is not a 1.5$\sigma$ significance, since we are considering two distributions of probability (the continuum and the photometric point). Combining both probability distributions and both thresholds we estimate that, if there was no emission, we would obtain less than 0.2\% false positives (assuming normal distributions for the photometric errors).

Five galaxies showing significant emission in two or more filters where no strong lines should be present (according to their redshift) were removed. In addition, visual inspection of the SEDs of limiting cases led to the removal of five objects and the addition of seven. The added galaxies correspond to cases where the continuum fit was artificially widened by one bad photometric point, lowering the significance of emission lines below the threshold. The removed galaxies showed a very noisy SED that was not automatically identified in previous rejection procedures. Morphological inspection of all selected galaxies resulted in the removal of two galaxies that were, in fact, star-forming regions of spiral galaxies, and one overlapping galaxy pair, where a background galaxy contaminated the ELG detected with our algorithm.

In addition, we cross-matched our catalog with that of \cite{2016ApJS..224...15X}, which gathers X-ray sources in the GOODS-N field. Six sources in the sample show X-ray emission within a three-arcsecond radius, one of which was identified as an active galactic nucleus (AGN), and it was therefore removed from the sample. Considering the sensitivity limits of the survey ($\sim 2 \times 10^{40}$ erg\ s$^{-1}$ at $z$=0.35, $\sim 7 \times 10^{39}$ erg\ s$^{-1}$ at $z$=0.2), we rule out AGN contamination except for some low-luminosity AGNs.

\section{Sample of ELGs}
\label{sample}
\begin{figure}
   \centering
   \includegraphics[width=0.5\textwidth,keepaspectratio]{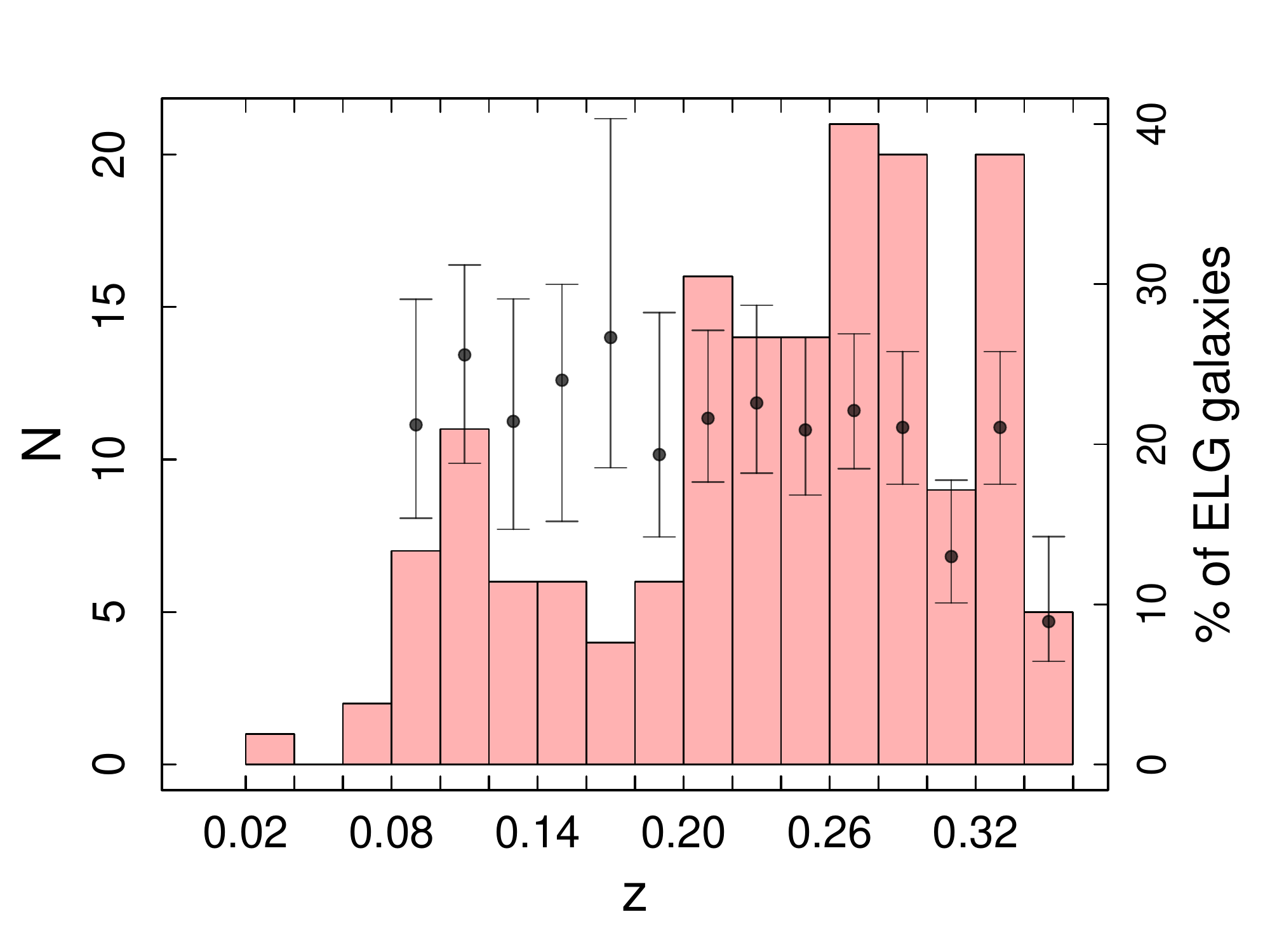}
      \caption{Distribution of redshifts of the ELG sample. For the galaxies that lack spectroscopic redshift determination, we use the value determined by our emission line detection algorithm. The gray dots represent the amount of ELG as a percentage of the reference sample for each redshift bin as shown in the right axis. }
         \label{fig:distredshift}
   \end{figure}
Our final sample consists of 160 ELGs obtained from the SHARDS survey in the GOODS-N field. 104 of them have spectroscopic redshift determination, from which 100 one-dimensional spectra are available: TKRS and DEEP-2 surveys provide 76, the rest being provided by S. Barger (private communication). Considering the two subsamples separately, we see that the subsample without spectroscopic observations presents a median apparent magnitude of 24.75 $\pm$ 0.13 (considering all SHARDS filters), $\sim$ 1.85 magnitudes fainter than the spectroscopic sample. This is due to the selection criteria for spectroscopic surveys, prioritizing brighter targets, and means that in this work we identify low-luminosity galaxies, unaccessible with previously available spectroscopic surveys. Moreover, 60 galaxies are detected in the ALHAMBRA survey, and 43 with GALEX. The main physical properties of our ELG sample are summarized in Table \ref{tab:main_table}.

The redshift distribution is shown in Fig. \ref{fig:distredshift}. The number of galaxies increases with redshift, as expected due to the higher volume of Universe considered. The sharp decline at $z$>0.34 corresponds to the wavelength limit for H$\alpha$ in SHARDS. The lower value between 0.3 and 0.32 does not correspond to any detection limit and seems to be barely significant when considering the density of galaxies at that redshift range in the parent sample. For comparison purposes we have defined a reference sample in SHARDS which we use throughout this paper. It consists of all galaxies in the catalog with $z$<0.36 and absolute magnitude in F850LP brighter than 90\% of the ELG sample (as a function of redshfit) totaling 779 galaxies. The original parent sample (all sources detected in SHARDS with $z<0.36$) was larger (1823 sources) but most of those galaxies are faint, and therefore comparing with them would be biased.

Figure \ref{fig:distredshift} also shows (with black dots) the amount of ELGs  as a percentage of the number of galaxies in the reference sample per redshift bin (error bars are computed following \citealt{2011PASA...28..128C}, here and in similar plots throughout this paper). The fraction of ELG is $\sim$ 22\% and remains constant within our redshift range.

\begin{table*}[t]
\fontsize{6.7}{9}\selectfont
  \centering
  \caption{List of ELGs identified in SHARDS and its main properties. EWs and H$\alpha$ luminosity are corrected by extinction. The complete table is available in the online version; only the first rows are shown here as guidance.}
  \label{tab:main_table}

  \begin{tabular}{ccccccccccccccc}
\hline\hline
\rule{0pt}{2ex}    ID & RA & DEC & $z_{phot}$ &$ z_{spec} $& $z_{lines} $& $ pEW_{H\alpha}$& $ sEW_{H\alpha}$&$ F_{H\alpha}$&  $pEW_{[OIII]}$&$ sEW_{[OIII]}$&$ F_{OIII}$ & $L_{H\alpha}$ & $F850LP$ & N2 \\
(1) & (2) & (3) & (4) & (5) & (6) & (7) & (8) & (9) & (10) & (11) & (12) & (13) & (14) & (15) \\

\hline
\rule{0pt}{2ex}   20002280 & 189.3239 & 62.19090 & 0.21 & 0.213 & 0.22 & 65 ± 14 & 88 ± 12 & 72 ± 11 & 34 ± 17 & 44 ± 4 & 60 ± 25 & 153 ± 23 & 22.661 ± 0.019 & 0.14\\ 
  10000098 & 189.3262 & 62.19741 & 0.11 & 0.105 & 0.11 & 19 ± 13 & 9.2 ± 0.4 & 169 ± 97 & 18 ± 12 & 3.4 ± 0.2 & 124 ± 71 & 53 ± 31 & 21.046 ± 0.002& 0.38 ± 0.04 \\
  10000145 & 189.3580 & 62.20181 & 0.09 & 0.089 & 0.09 & 45 ± 8 & 55 ± 1 & 182 ± 26 & 26 ± 10 & 29 ± 1 & 154 ± 48 & 57 ± 8 & 21.313 ± 0.006 & 0.05 \\ 
  10000515 & 189.3600 & 62.22278 & 0.30 & 0.299 & 0.29 & 43 ± 28 & 21 ± 2 & 26 ± 13 & 16 ± 8 & 11 ± 2 & 17 ± 8 & 181 ± 89 & 23.257 ± 0.022 & 0.48 \\ 
  10000777 & 189.3220 & 62.23236 & 0.33 & 0.336 & 0.33 & 32 ± 17 & 41 ± 2 & 194 ± 89 & 20 ± 8 & 21 ± 1 & 177 ± 57 & 1078 ± 492 & 20.971 ± 0.002 &  0.13\\ 

   & & & & & & ..... & & &  \\
  \hline 
\multicolumn{12}{l}{\rule{0pt}{4ex}(1) SHARDS ID.}\\
\multicolumn{12}{l}{(2) Right ascension from SHARDS catalog.}\\
\multicolumn{12}{l}{(3) Declination from SHARDS catalog.}\\
\multicolumn{12}{l}{(4) Photometric redshift from SHARDS catalog.}\\
\multicolumn{12}{l}{(5) Spectroscopic redshift from \cite{2008ApJ...689..687B}.}\\
\multicolumn{12}{l}{(6) Photometric redshift derived in this study using H$\alpha$ and [OIII]5007 lines.}\\
\multicolumn{12}{l}{(7) Photometrically derived H$\alpha$ EW, in \text{\AA}ngstroms.}\\
\multicolumn{12}{l}{(8) Spectroscopically derived H$\alpha$ EW, in \text{\AA}ngstroms.}\\
\multicolumn{12}{l}{(9) Photometrically derived H$\alpha$ flux, in $10^{-18}$ erg\ s$^{-1}$\ cm$^{-2}$.}\\
\multicolumn{12}{l}{(10) Photometrically derived [OIII]5007 EW, in \text{\AA}ngstroms.}\\
\multicolumn{12}{l}{(11) Spectroscopically derived [OIII]5007 EW, in \text{\AA}ngstroms.}\\
\multicolumn{12}{l}{(12) Photometrically derived [OIII]5007 flux, in $10^{-18}$ erg\ s$^{-1}$\ cm$^{-2}$.}\\
\multicolumn{12}{l}{(13) Photometrically derived H$\alpha$ luminosity, in $10^{38}$ erg \ s$^{-1}$.}\\
\multicolumn{12}{l}{(14) Magnitude of the galaxy in F850LP band, from \textit{HST} \citep{2014ApJS..214...24S}.}\\
\multicolumn{12}{l}{(15) [NII6583]/H$\alpha$ ratio derived from spectroscopy. If no uncertainty is shown, the value is an upper limit.}\\
\end{tabular}
\end{table*}

Our main motivation for performing a blind search to find emission lines was the uncertainty in the photometric redshifts. Our detection procedure, based on identifying both H$\alpha$ and [OIII]5007 lines allows for a more precise redshift determination compared to the previous photometric redshift. Even considering the high accuracy of SHARDS photometric redshifts ($\frac{\Delta z}{z} \sim 0.0055$), small changes in redshift would change the filter where we expect to detect the emission lines. Using the methodology described in this study, we recover 22 galaxies which would have been mistaken as non-emitters if we had relied in the previous photometric redshifts only. We marginally improve the redshift accuracy (up to  $\frac{\Delta z}{z} \sim 0.0035$).

\subsection{Equivalent widths and line fluxes}
\label{EWcalc}

      \begin{figure}
   \centering
   \includegraphics[width=0.5\textwidth,keepaspectratio]{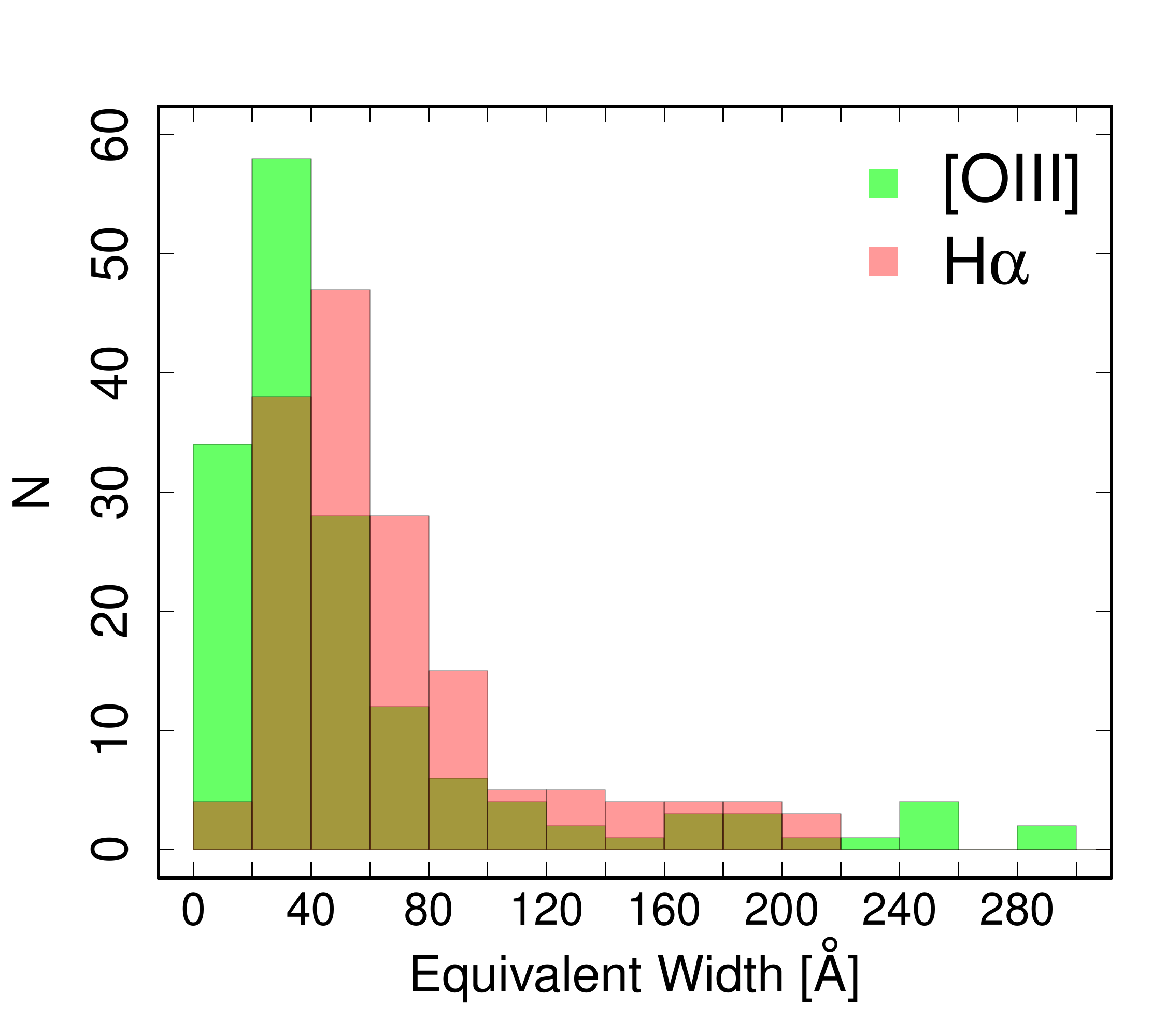}
      \caption{Distribution of H$\alpha$ and [OIII]5007 EWs in the sample.}
         \label{fig:EW}
   \end{figure} 

A key parameter providing insight into the characteristics of the stellar population(s) of each galaxy  is the EW in H$\alpha$. This parameter carries valuable information on the age and strength of the star-forming burst, and we use it as a further constraint in our stellar population modeling of the SED (see Section \ref{SEDfit}).

For the derivation of the EW it is necessary to estimate the flux of the stellar continuum underneath the line. In order to do so, we performed a weighted median of the flux in the two or three filters lying blue-ward of H$\alpha$ (red-ward for the [OIII]5007 line). If the filter adjacent to the line was contaminated by it, only two filters were used. The median was performed with a weighted Bootstrap and Monte Carlo method over those two or three filters, giving more weight to the filters closer to the line. This method for estimating the continuum was appropriate due to the minimum slope in the continuum around the lines (in $\mu$Jy). It was preferred to a linear fit to several filters on both sides of the lines because, red-ward of H$\alpha$ (blue-ward of [OIII]5007), there might be a strong contamination of [NII] and [SII] ([OIII]4959 and H$\beta$). Furthermore, in several galaxies, the spectral region red-ward of H$\alpha$ is covered by the wider SHARDS filters (F883W35 and F941W33, with FWHM $\sim$ 300 \text{\AA}) leading to higher uncertainty. Using linear extrapolations from only one side of the line would artificially increase uncertainties, and using the continuum derived in subsection \ref{em_lin_detec} would add constraints from distant regions of the spectrum that could bias the estimation. Comparing these values with those obtained using the continuum derived in the SED fitting (section \ref{SEDfit}) shows no bias and a small scatter.

After we estimate the flux density of the continuum, $F^{\lambda}_{cont}$, we subtract it from the flux density in the filter where the line lies ($F^{\lambda}_{Fl}$) to obtain the line flux $F_{line}$ and compute the EW as: 
\begin{equation}
EW=\frac{(F^{\lambda}_{Fl}-F^{\lambda}_{cont})\times\Delta}{F^{\lambda}_{cont}}=\frac{F_{line}}{F^{\lambda}_{cont}}
,\end{equation} 
where $\Delta$ represents the width of the filter. The error in the EW is computed propagating the error in the photometric value of the filter where the line lines and the error in the continuum (taking the 68\% central values of the weighted Bootstrap and Monte Carlo simulations).

Using the subsample with available spectra, we can evaluate the contamination of the measured EW by other lines. In 53\% of the galaxies, the filter where we detect [OIII]5007 is contaminated by the [OIII]4959 line, but in no case is it affected by H$\beta$. When present, this contamination (as measured in galaxies with available spectra), is $\sim$ 35\% of the value of [OIII]5007, very similar to the one derived from theoretical models ($\sim$ 34\%, from \citealt{2000MNRAS.312..813S}). When considering the reported values of [OIII]5007 EW this caveat should be taken into account.

Regarding the filter that corresponds to H$\alpha$, it is contaminated by the [NII]6583 line in 86\% of the galaxies. The average contamination by this line is $\sim$ 15 \% of the value of H$\alpha$ (measured in the available spectra). Given the uncertainties in the photometric EW determination, correcting for this effect could be problematic. For example, if the H$\alpha$ line falls in a wavelength where the filter transmission is lower than 100\%, we would be underestimating its actual EW. Considering the spectra, this effect would be of 10\% on average, which nearly offsets the effect of [NII]. We therefore make no correction to the measured values, but we take the uncertainty into account when discussing the results.

For the galaxies where H$\alpha$ is detected in the reddest filter (F883W35), [SII] lines contaminate H$\alpha$ in 73\% of the cases, and their flux accounts for 30\% of H$\alpha$ flux on average. We corrected it by this amount in the galaxies where the [SII] lines match the filter wavelength range. In summary, we are aware of the possible contamination of the H$\alpha$ and [OIII]5007 fluxes, and we correct for it in the particular case of [SII] lines. 

The distribution of the EWs both in H$\alpha$ and [OIII]5007 emission lines is shown in Fig. \ref{fig:EW}.

\subsection{Completeness of the sample}

\begin{figure}

   \centering
   \subfigure{ \includegraphics[width=0.47\textwidth,keepaspectratio]{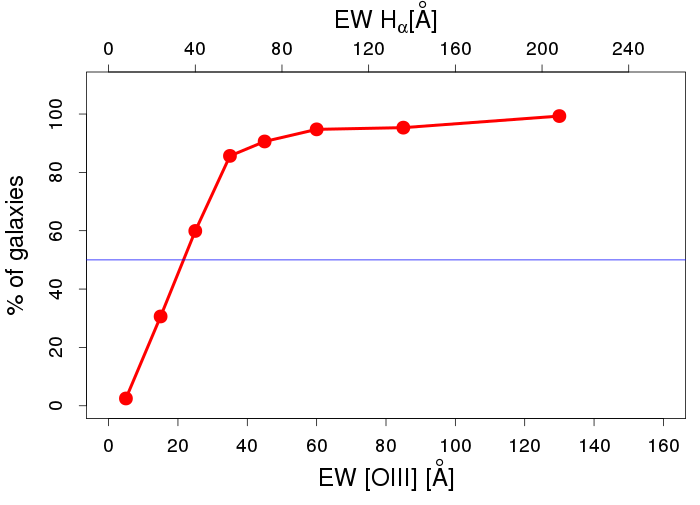}}
   \subfigure{ \includegraphics[width=0.47\textwidth,keepaspectratio]{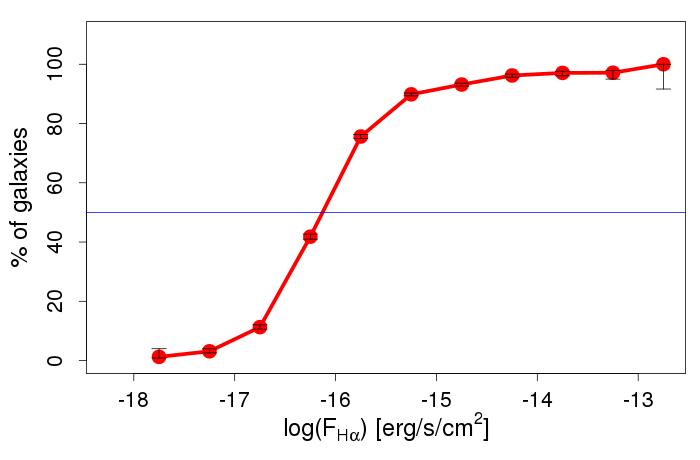}}
      \caption{Completeness of the sample, derived from simulated SEDs with emission lines added. \textit{Top:} Completeness as a function of the EW of \lbrack OIII\rbrack \ and H$\alpha$ lines. We reach 50\% completeness at around 22 {\AA} in [OIII], 35 {\AA} in H$\alpha$. \textit{Bottom:} Completeness of the sample, using the same simulations as in the top panel, but plotted as a function of the H$\alpha$ flux. We reach 50\% completeness at around 7.4\ 10$^{-17}$ erg\ s$^{-1}$\ cm$^{-2}$ }
         \label{fig:simul_complet}
   \end{figure}
   
To estimate the completeness of the ELG sample and the limitations of our detection procedure, we run a series of simulations. We create synthetic SEDs and feed them into our algorithm to compute the percentage of detections as a function of the EW and line flux.

We use the synthetic spectra that we obtain in Section \ref{SEDfit} with SED fitting techniques, and we add [OIII]5007 and H$\alpha$ emission lines of different EW, from 0 to 150 \text{\AA}. H$\alpha$ EW was set 1.6 times larger than [OIII]5007 EW (as it is the median ratio in the observed sample). We also took into account the shift in central wavelength of each filter depending on the position of the galaxy in the field of view (see \citealt{2013ApJ...762...46P} for more details) choosing a random set of shifts for each simulated galaxy. Then, the spectra were convolved with the SHARDS filters, and the Monte Carlo method was applied to them, using uncertainties similar to those present in the parent catalog.

We then run our detection code on each simulated galaxy (for a total of 20280). The percentage of successful detections is shown in Fig. \ref{fig:simul_complet} as a function of the EW and H$\alpha$ flux. We reach 50\% completeness at around 22 {\AA} in [OIII]5007, 35 {\AA} in H$\alpha$ and $\sim 10^{-16} $ erg\ s$^{-1}$\ cm$^{-2}$ in H$\alpha$ flux. This result is consistent with the properties of the detected sample (see Fig. \ref{fig:EW}): the number of galaxies grows with decreasing EW values, down to $\sim$ 20 {\AA} in [OIII]5007 and $\sim$ 40 {\AA} in H$\alpha$, where it starts decreasing.

We reach limits, both in EW (min. $\sim$ 15 \text{\AA}) and flux (median $\sim$ 4$\times$ 10$^{-17}$ erg\ s$^{-1}$\ cm$^{-2}$), that are similar to those found in \cite{2015ApJ...812..155C}, and comparable to those of narrow-band surveys \citep{2013MNRAS.428.1128S} with a much wider redshift coverage.
To compare our detection efficiency with \cite{2016A&A...592A.122H}, we consider only galaxies in our sample with EW>80 {\AA} in both lines; we find that those ELG make up 2.5\% of our reference sample, compared to the value of $\sim$ 1\% they find. Considering only galaxies with spectral coverage, these values grow to 6\% and 3\%, respectively. We detect approximately two times more galaxies, probably due to the improved depth and wavelength coverage of the SHARDS survey. 

In order to further investigate the completeness of the sample, we check how many galaxies with emission lines in the spectra are recovered with our code. Considering all spectra with EW in both lines over the 50\% completeness threshold, we find 57 galaxies. Out of these, 8 (14\%) are not detected by our code in SHARDS photometry due to insufficient S/N in the lines (in most cases, due to one of the lines falling in a gap between filters). All galaxies where the spectral EW fulfills the threshold for 80\% completeness are detected with our code. These results are better than what the simulations predict, but they are consistent with the higher luminosity of targets with spectra available, which makes it easier to spot emission lines in the photometry.

\section{SED fitting and models}
\label{SEDfit}

To unveil the physical characteristics of the sample of ELGs, we performed stellar population fitting to their SED with our own taylor-made code for this work. Our main assumption was to model the galaxies with two single stellar population (SSP) models: a young instantaneous burst for the star-forming component, and an old burst for the underlying host galaxy. We create a library of SSP models using \textit{Starburst99}\footnote{\url{http://www.stsci.edu/science/starburst99/docs/default.htm}} software \citep{1999ApJS..123....3L, 2014ApJS..212...14L} for instantaneous SF, with stellar mass normalized to $10^6 M_{\odot}$. We used a \cite{1955ApJ...121..161S} initial mass function (IMF) ($\alpha=2.3$) between 0.1 and 120 $M_{\odot}$, with the standard Geneva evolutionary tracks \citep[][and references therein]{1999A&AS..135..405C}. 

The ranges of variation in the SSP parameters considered are the following:
\begin{itemize}
\item{\textbf{Metallicity of the young population}}: Z = 0.001, 0.004, 0.008, and 0.02.
\item{\textbf{Extinction of the young population}}: E(B-V) from 0 to 0.5, with steps of 0.04 (13 values). It corresponds to $A_V$ from 0 to 1.55 mag.
\item{\textbf{Age of the young population}}: From 2.5 to 13 Myr, with steps of 0.5 Myr.
\item{\textbf{Metallicity of the old population}}: Fixed at Z= 0.004.
\item{\textbf{Extinction of the old population}}: E(B-V) fixed at 0.08.
\item{\textbf{Age of the old population}}: Fixed at 2 Gyr.
\item{\textbf{Burst strength}}: It was considered fixed for each combination of models (see item 2 in the current section).

\end{itemize}

Once the observed SEDs and the SSP models were defined, we performed the fitting procedure as follows.

\begin{enumerate}

\item Given the set of ages and metallicities for the young and old stellar populations, every possible combination of the parameter space was considered.

\item For each combination, we compute the burst strength (the mass ratio between the young and old populations) that produces the observed EW in H$\alpha$. The key in our analysis is the use of the H$\alpha$ EW measured photometrically (Sect. \ref{EWcalc}). Given two SSPs, only one ratio between them results in the observed H$\alpha$ EW. Combining both populations accounting for this factor we obtain the master composite stellar populations (CSP).

Starburst99 models provide as output the luminosity in H$\alpha$, assuming case-B recombination, where all Lyman continuum photons are reabsorbed. The model EW is computed as: 
\begin{equation}
EW(H\alpha)_{model}=\frac{F_{line}(H\alpha)}{B_R \times F_{cont.old}[6563 \text{\AA}]+F_{cont.you}[6563 \text{\AA}]}
,\end{equation}

where $F_{cont.old}[6563 \text{\AA}]$ and $F_{cont.you}[6563 \text{\AA}]$ are the flux densities of the continuum of the old and young populations, respectively, at the wavelength of H$\alpha$. $B_R$ is the flux (and mass) ratio between the old and the young populations.

\item For each CSP model, we apply the set of extinction corrections, using the extinction law derived by \cite{2003ApJ...594..279G} for the bar of the Small Magellanic Cloud, with $R_V$=3.1. We then compute the median ratio between the model photometry and the observed photometry to derive the mass, allowing for a $ \pm 2\%$ variation. To derive the right extinction and mass for each CSP, we fit it to the photometry using a $\chi^2$ minimization: 
\begin{equation}
\chi^2_{model}=\frac{\sum_i[(F_{model,i}-F_{obs,i})/\Delta F_{obs,i}]^2}{n_{filters}-n_{param}}
,\end{equation}

where $F_{model,i}$ is the flux of the model in each of the $i$ filters used, $F_{obs,i}$  the observed one and $\Delta F_{model,i}$  its error. $n_{filters}$ is the number of $i$ filters used in the fit and $n_{param}$ is the number of free parameters in the models.

\item Finally, having the best extinction and mass for each CSP model, we select the one that minimizes the $\chi^2$.

\end{enumerate}

Four examples of SED fitted galaxies are shown in Fig. \ref{fig:ejemplos_SEDS}, with different weights between the old and young stellar populations and extinction. Red and blue lines correspond to the synthetic spectra of the old and young stellar populations, respectively, while the dark green line represents the sum of the two. 

The error on each parameter of the fit was computed using Monte Carlo simulations. For each galaxy, we generated 600 realisations of the SED allowing each photometric point to vary within its photometric error. Then, we ran the code on each simulated SED and compiled the distribution of values for each fitted parameter. We considered the width of this distribution (containing 68\% of the simulations) as the uncertainty in the parameter, and they are shown in Table \ref{tab:sed_table}. The computed error are asymmetric and sometimes the best fitted parameter represents either a lower or an upper limit of the distribution. This is mainly due to both the coarse possibilities for the metallicity values and the correlation between metallicity and age (an thus extinction and mass).

\subsection{Input of the SED-fitting procedure}
\subsubsection{Completing the SED with ALHAMBRA and GALEX}
\label{rescale}

      \begin{figure}
   \centering
   \includegraphics[width=0.5\textwidth,keepaspectratio]{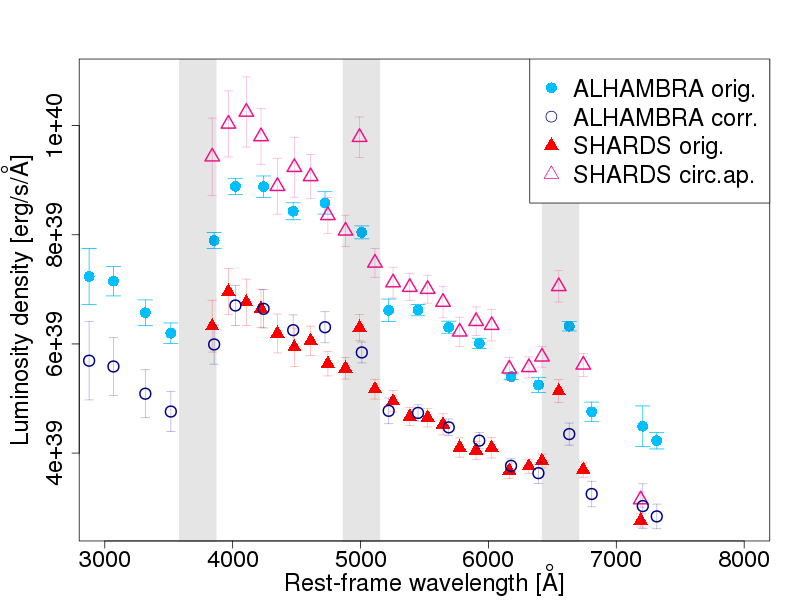}
      \caption{SED of the galaxy SHARDS10001384. SHARDS photometry (red triangles) and ALHAMBRA (filled blue circles). The pink open triangles represent the interpolation between the values derived from the circular SHARDS apertures encompassing ALHAMBRA, and the open dark blue circles are the result of applying the scaling factor to ALHAMBRA data. We see that the values obtained using larger SHARDS apertures and those from ALHAMBRA are similar. Shaded regions are wavelength ranges contaminated by prominent emission lines and are not used in the derivation of the scaling factor.}
         \label{fig:fotometria_al}
   \end{figure}

   \begin{figure}
   \centering
   \includegraphics[width=0.5\textwidth,keepaspectratio]{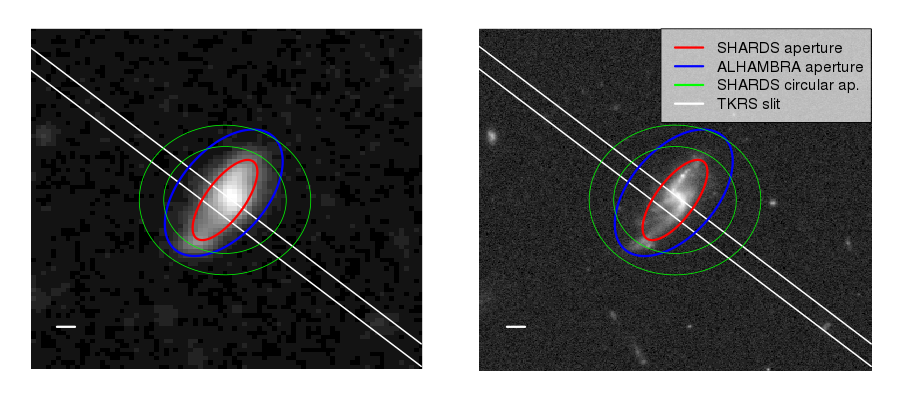}
      \caption{Postage stamp of a galaxy (SHARDS10001384) as seen in a SHARDS image (left) and a HST-ACS one (right). We overplot the SHARDS aperture (red), the ALHAMBRA aperture (blue) and two SHARDS circular apertures encompassing ALHAMBRA. We also represent in white the slit used by the TKRS survey to obtain the long-slit spectrum. The white line at the bottom-left corner is one arcsecond long.}
         \label{fig:imagenes_aperturas}
   \end{figure}

   \begin{figure}[!htbp]
   \centering
   \subfigure{ \includegraphics[width=0.48\textwidth,keepaspectratio]{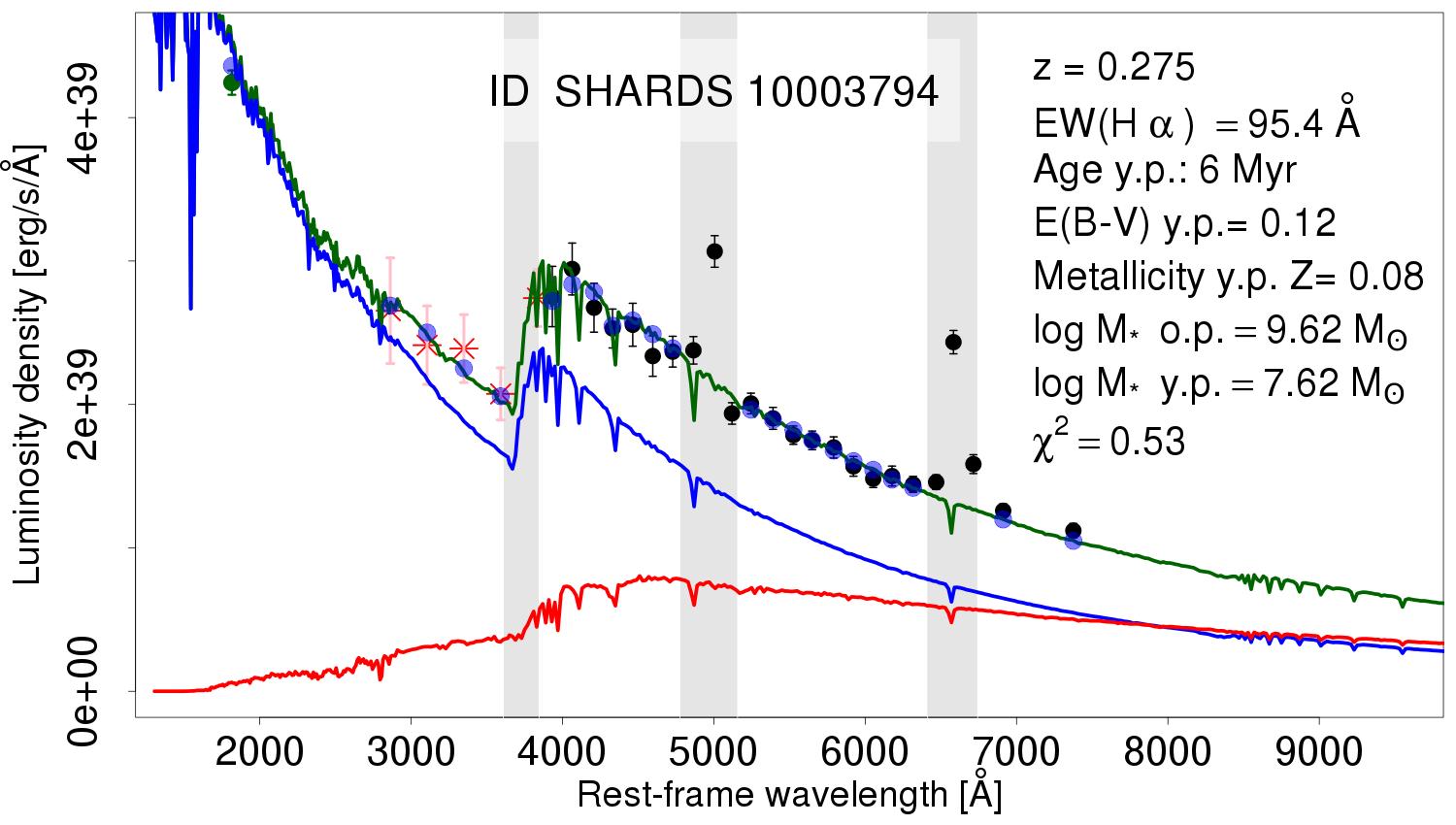}}
   \subfigure{ \includegraphics[width=0.48\textwidth,keepaspectratio]{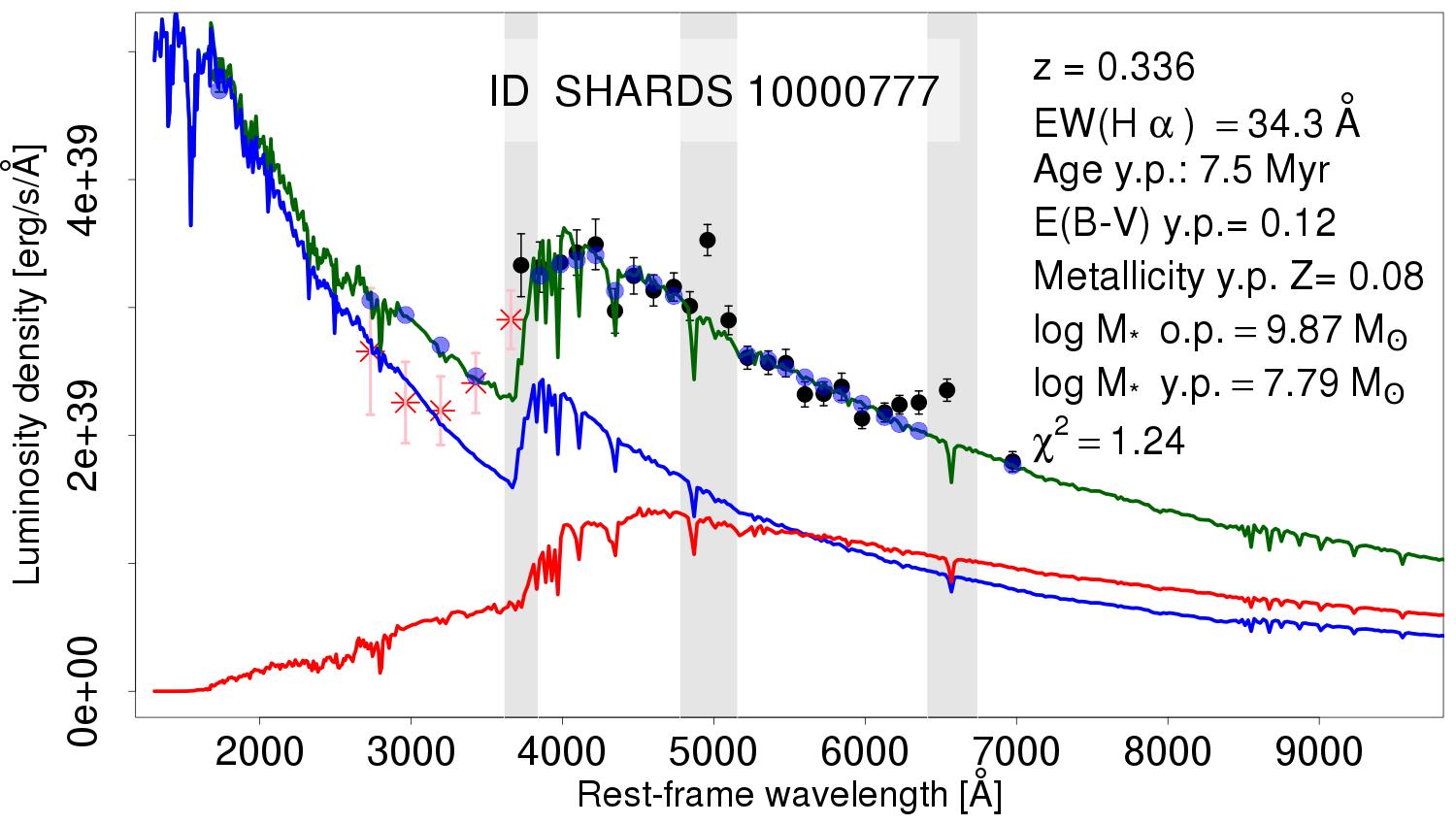}}
   \subfigure{ \includegraphics[width=0.48\textwidth,keepaspectratio]{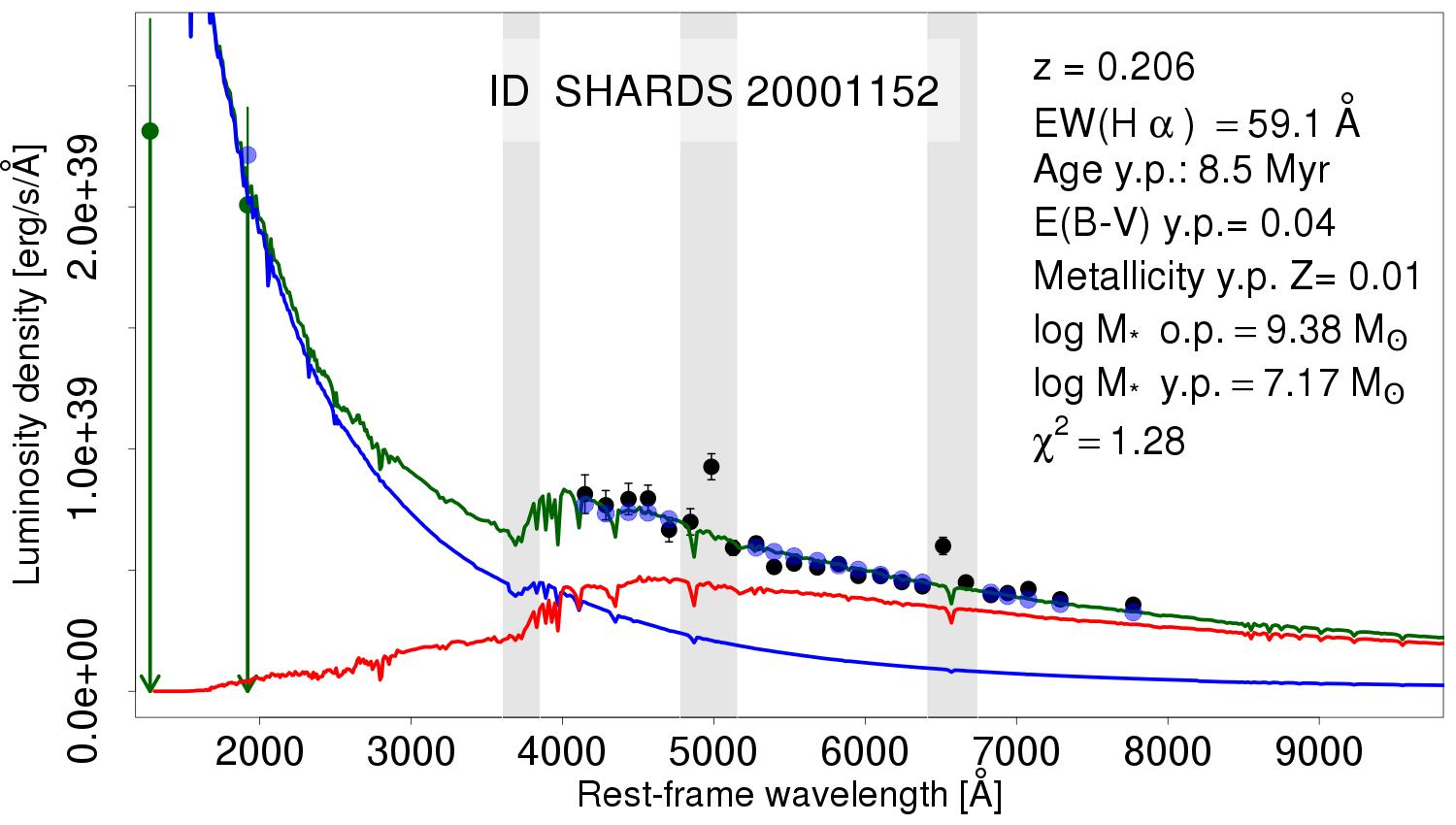}}
   \subfigure{ \includegraphics[width=0.48\textwidth,keepaspectratio]{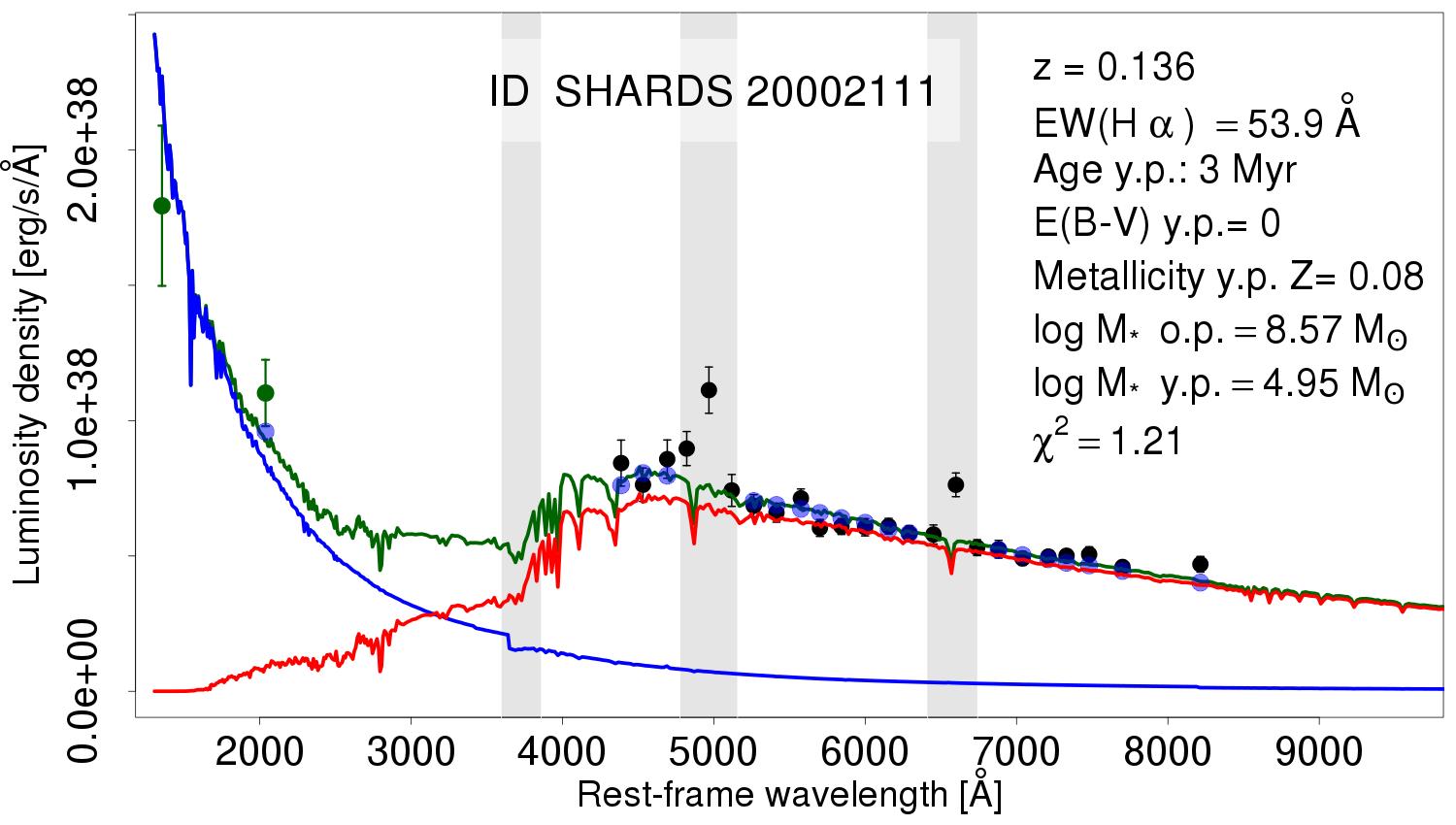}}
\caption{Example of four SED fitted galaxies, with a range of different physical parameters, shown in the top right corner. Black dots correspond to SHARDS photometric points, red stars to ALHAMBRA data (when available), and dark green dots to GALEX data (only upper limits in the third panel). The red line represents the stellar spectrum of the old stellar population, while the blue one represents the young population, and the sum of both is shown in dark green. Gray shaded areas cover the filters that could be contaminated with nebular emission lines, and are not taken into account when deriving the $\chi^2$ value.}
         \label{fig:ejemplos_SEDS}
   \end{figure}

When comparing ALHAMBRA and SHARDS photometry, a small offset between them was noticeable (Fig. \ref{fig:fotometria_al}), caused by either an offset in the absolute photometric calibration of the surveys, or by aperture differences. When measuring SHARDS photometry in circular apertures with similar radius to the one used in ALHAMBRA, the offset mostly disappeared, which indicates the issue was only due to aperture sizes (see Figs. \ref{fig:fotometria_al} and \ref{fig:imagenes_aperturas}). For 65\% of the sample, ALHAMBRA aperture is larger, with a median of 0.43" larger semi-major axis (the remaining 35\% are 0.18" smaller in median).

In order to correct for this aperture difference, we compute an empirical scaling factor between both surveys (since ALHAMBRA images are not yet publicly available). For each galaxy, we calculated the ratio between every SHARDS photometric point and the closest one in ALHAMBRA, and then fitted a linear model (using the Bootstrap and Monte Carlo methods) to those ratios, as a function of wavelength. We extrapolated the linear fit to bluer wavelengths to obtain the factor for each one of the ALHAMBRA filters, and used the width of the distribution as uncertainty. The median value of the scaling factor is $\sim$ 1.2. 

In order to take into account the spectral region bluer than H$\alpha$, we added to the fit only the ALHAMBRA photometric points with shorter central wavelengths than the bluest SHARDS filter. Adding all the ALHAMBRA points would have overstressed the importance of fitting the redder wavelengths of the galaxy, when in fact more information about the age and extinction is stored in the bluest ones (where the Balmer break lies).

We also include in our SEDs GALEX photometry for both FUV and NUV bands. Given the large PSF ($\sim 6"$) of GALEX data, contamination by other sources was often present, making it necessary to visually inspect all detected galaxies to identify those affected by contamination. In those cases (as well as those where the galaxy was not detected in GALEX images) we considered GALEX data only as an upper limit in the SED fits. FUV photometry was only used when the rest-frame wavelength range covered on the galaxy was larger than 1100 \text{\AA}, since at lower values the extinction law used in this work becomes very uncertain.

\subsection{Tests on the input parameters of the models}

\subsubsection{Age of the old stellar population} \label{oldpop}

   \begin{figure}
   \centering
   \includegraphics[width=0.5\textwidth,keepaspectratio]{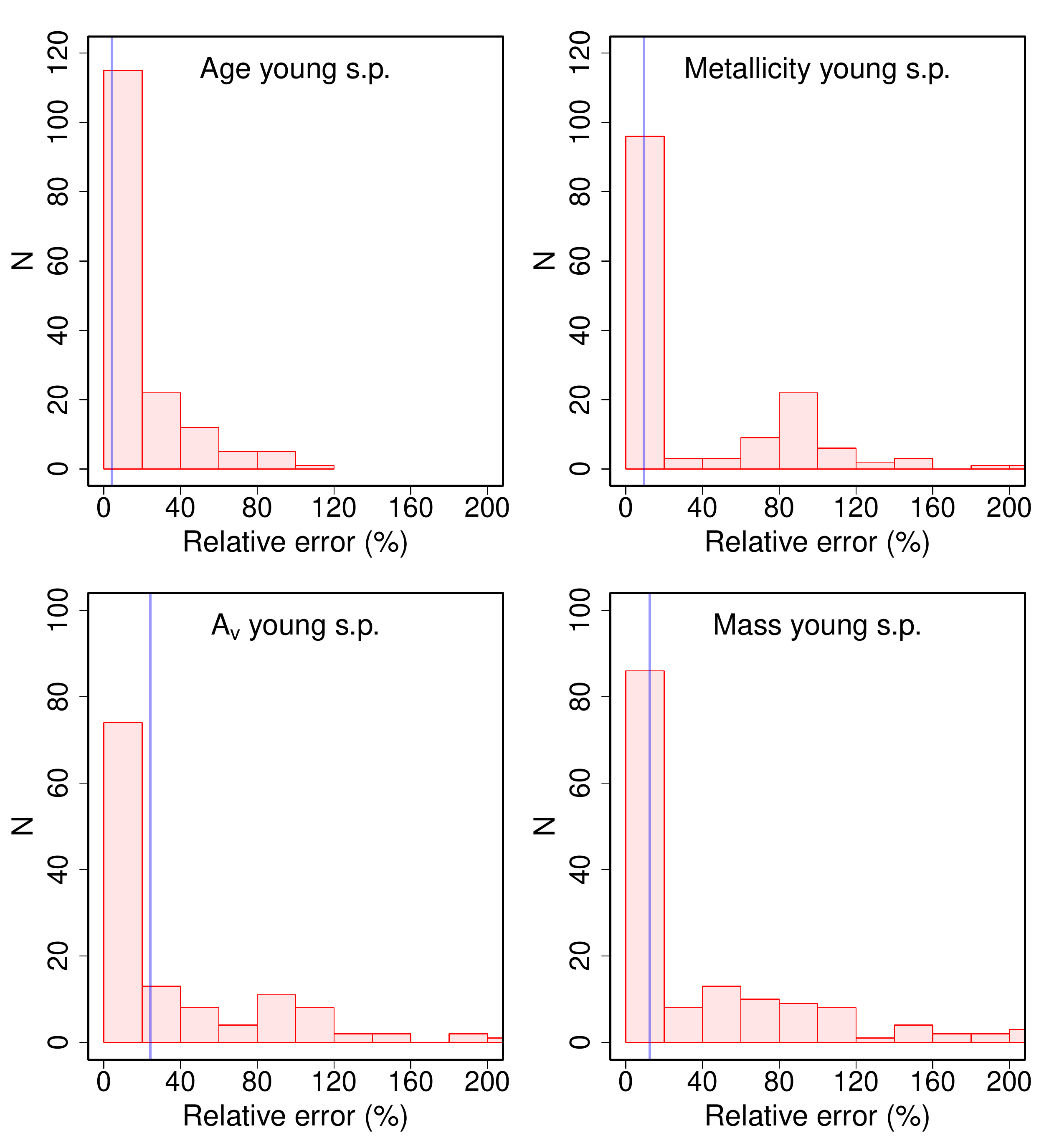}
      \caption{Histograms showing the relative uncertainty in each parameter of the young stellar population caused by the degeneracy in the old population. Panel (a) represents age, (b) metallicity, (c) extinction ($A_V$) and (d) mass. We derive the best value for each parameter in each galaxy, for all considered old stellar population ages. Then we take the width of its distribution for each galaxy and compute the relative error. We see that the changes in the best values are smaller than 20\% in most cases. The blue lines represent the median value of the relative error.}
         \label{fig:dispersiones}
   \end{figure}

When running the SED fitting code using the age of the old stellar population as a free parameter, we noticed a large degeneracy among models from 1 to 8 Gyr, where the $\chi^2$ values varied very little. This is caused by the small difference in the shape of the synthetic spectrum for populations spanning this range of ages: the most prominent change being simply the global loss of flux as the population grows older. However, this is easily compensated with a smaller extinction and higher mass, so we can get almost the same spectrum for different ages adjusting the other parameters.

We have studied the effects of using different values of the age for the old stellar population on the remaining free parameters of our analysis. For each galaxy, we performed the fit using eight different ages for the old stellar population. Figure \ref{fig:dispersiones} shows the relative difference in the young stellar population's free parameters with respect to the median value. For each galaxy we obtain eight values for each parameter, measure the width of this parameter distribution, and compute the relative error, which is then presented in Fig. \ref{fig:dispersiones}. 

Most galaxies show very small dispersion in age, metallicity, mass, and extinction of the young stellar populations (< 20 \% of the median value for each parameter in each galaxy). Median dispersions are 0.5 Myr, 0.00375 dex, $1.6\times10^{6} M_{\odot}$, and 0.07 mag in age, metallicity, mass, and extinction (A$_V$) of the young stellar population, respectively. Therefore, we are confident in the values we compute for the parameters of the young stellar population. 

In order to obtain a set of reasonable fits, we chose to fix the age of the old population to 2 Gyr, following results by \cite{2016A&A...592A.122H} over a sample of similar characteristics. 

\subsubsection{Extinction of the old stellar population}
\label{extoldpop}
When performing tests where the old stellar population extinction was left free (with the same range as the young population), it was fitted to the most extreme values, acting as a degenerated parameter (increasing or decreasing the mass of the old population). As a result, we decided to fix it at E(B-V)=0.08 ($A_V$=0.25 mag).

\subsubsection{Metallicity of the old stellar population}
\label{metoldpop}

Considering the mass-metallicity relations presented in \cite{2014ApJ...791L..16G}, \cite{2015MNRAS.453.2490T} and \cite{2008MNRAS.391.1117P}, and the mass range of our sample, we chose to use a fixed $Z=0.004$ metallicity for the old component. We checked that using a higher metallicity ($Z=0.02$) produced some small scatter in most parameters of the fit, a $\sim$ 0.9 Myr increase in the age of the young populations, and a 0.2 dex decrease in mass.

\section{Spectra}
\label{spectra}

We made use of the available spectroscopic surveys to perform additional analysis to further characterize the properties of the 100 galaxies in the ELG sample with spectroscopic coverage.

\begin{figure}
   \centering
   \subfigure{\includegraphics[width=0.24\textwidth,keepaspectratio]{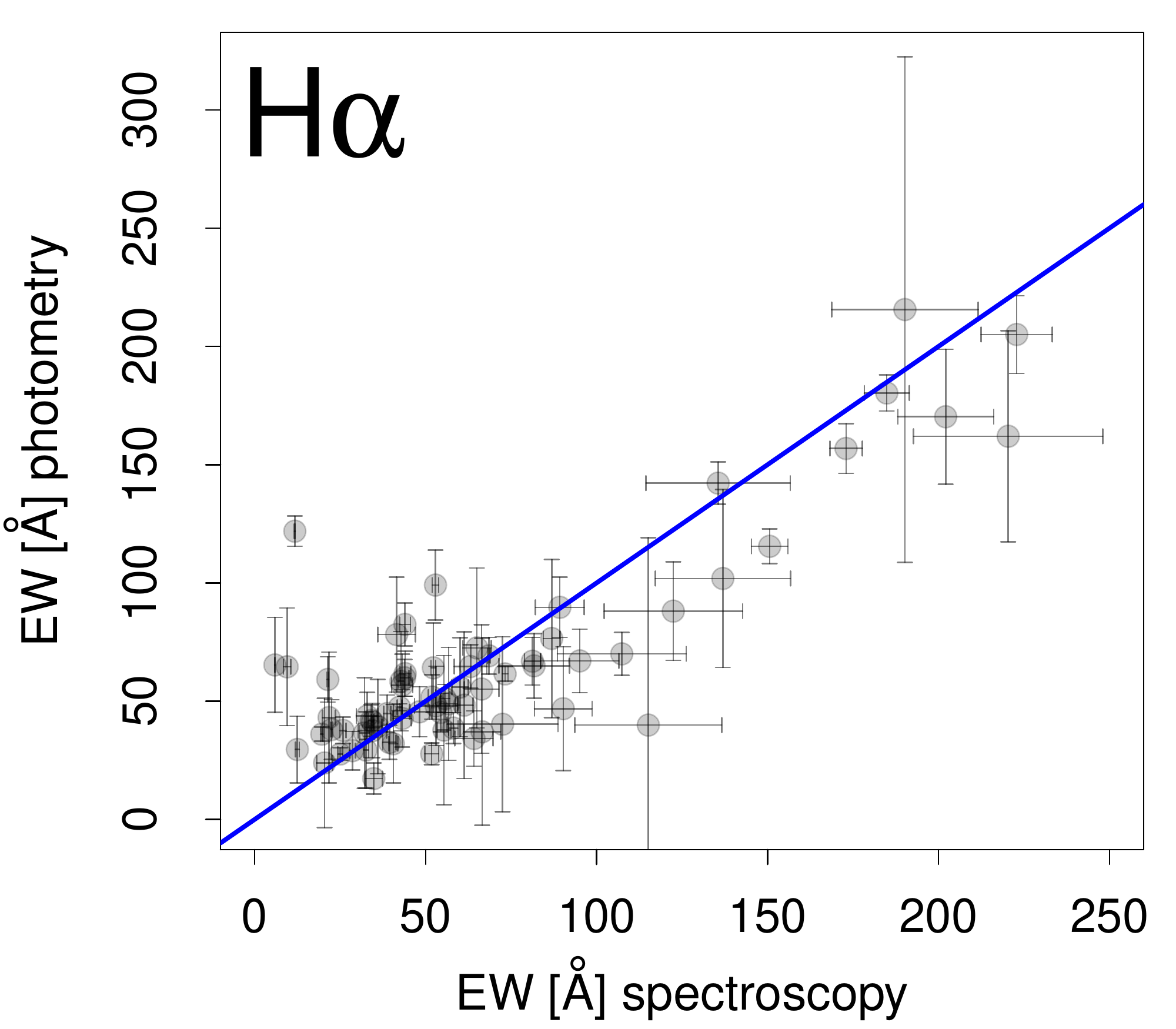}}
   \subfigure{\includegraphics[width=0.24\textwidth,keepaspectratio]{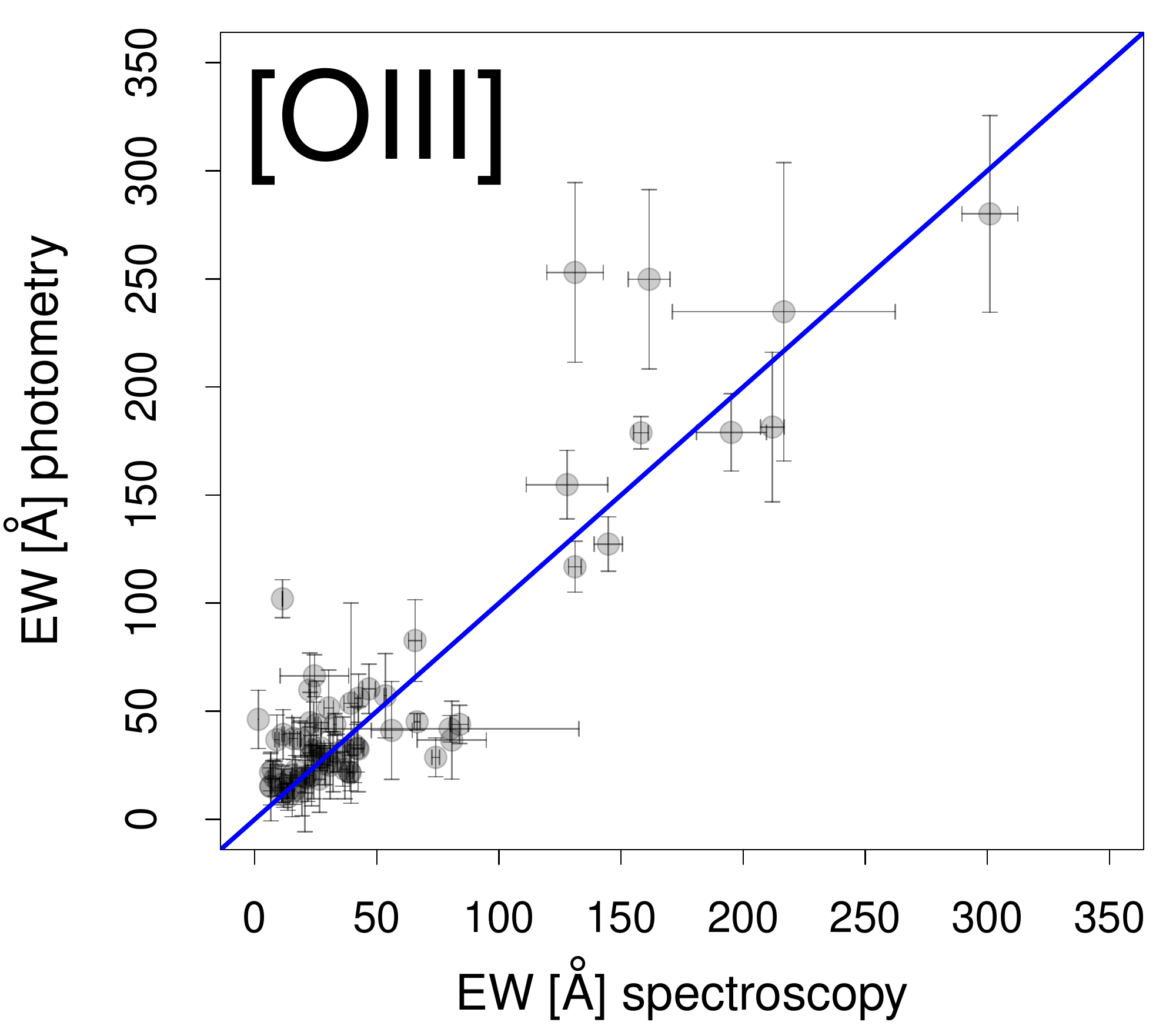}}
      \caption{Comparison between photometric EW and spectroscopic EW for H$\alpha$ (left panel) and [OIII]5007 (right panel). The blue line represents the one-to-one relationship.}
         \label{fig:EWo3}
   \end{figure}
In order to analyze the spectra we first apply a Gaussian smoothing to increase the S/N. Considering the instrumental width of the lines (1.4 \AA) and the width of the Gaussian smoothing function (1.7 \AA), the resulting spectral resolution is $\sim$ 2.2 \AA\. We calculate the S/N for each line measuring its peak value and dividing by the continuum noise at both sides of the line. In adding simulated lines, we compute a minimum equivalent width for each one to be detected, considering a S/N threshold value of 3. This enables us to set upper limits for the EW when the lines are not properly detected. The median S/N in the H$\alpha$ line is 27.

To derive the EW of the lines, a robust estimation for the continuum around each line was computed, fitting a linear model to the data points within a $RMSE$ around the median, at both sides of the line using Bootstrap simulations. Then, we computed the EW in different spectral apertures (ranging from 20 to 30 \AA), and combined the variation of the output value with the error in the continuum to produce the final uncertainty of the EW. The comparison between SHARDS photometric EW and spectroscopic EW measurements for both H$\alpha$ and [OIII]5007 lines is shown in Fig. \ref{fig:EWo3}. The deviation of the photometric and spectroscopic EW values for H$\alpha$ and [OIII]5007 are lower than 1$\sigma$ for 72\% and 68\% of the galaxies, respectively. These values raise to 97\% and 95\% for deviations lower than 3$\sigma$. Outliers are mainly due to misplacements of the slit and contamination of SHARDS filters with other lines (see Section \ref{EWcalc}).

We also use the emission line measurements to rule out AGN contamination in the sample. We measure the EW in H$\alpha$, [OIII]5007, [NII]6583 and H$\beta$ in the 26 galaxies where the S/N>3 in all lines, in order to build a BPT diagram \citep{1981PASP...93....5B}. For 40 extra galaxies we were only able to use the \cite{2011ApJ...736..104J} method (a [OIII]5007/H$\beta$ vs. mass diagram). After considering H$\beta$ absorption and poorly corrected sky lines at [NII] wavelength, we find no galaxy out of the star-forming region in either diagram. We therefore rule out AGN contamination in those 66 galaxies, on top of the X-ray data we used at the end of Section \ref{detection_lines}

We also estimate the metallicity of the gas using the empirical calibration based on the [NII]/H$\alpha$ ratio \citep{2002MNRAS.330...69D}. This method has been successfully applied to different samples of star-forming galaxies \citep{2011ApJ...743...77M,2015ApJ...810L..15S}. We use the calibration by \cite{2013A&A...559A.114M},

\begin{equation}
12 + \log(O/H) = 8.743 + 0.462 \cdot \log([NII]6583/H\alpha).,\end{equation}

\begin{figure}
   \centering
   \includegraphics[width=0.5\textwidth,keepaspectratio]{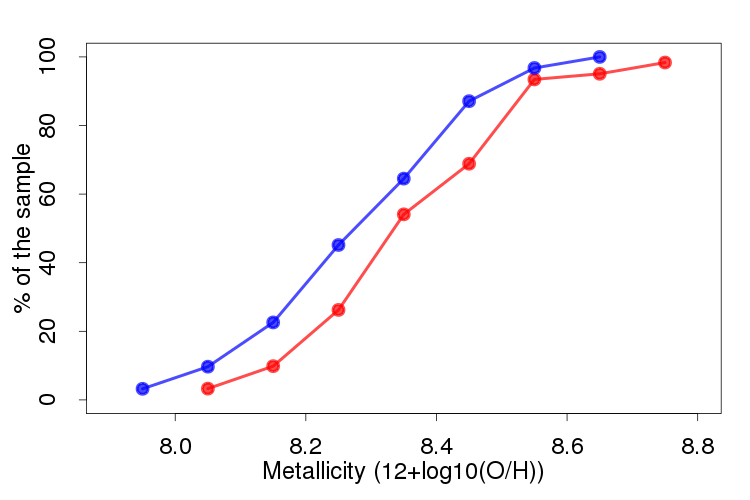}
      \caption{Cumulative distribution of gas metallicity of the galaxies in the spectroscopic sample. In blue, galaxies with detected [NII]6583 and thus measured metallicity. In red, those with only an upper limit to [NII]6583 and metallicity. }
         \label{fig:met_gas_cumul}
   \end{figure}

The results are shown in Fig. \ref{fig:met_gas_cumul}. All the ELG with spectra and good S/N show sub-solar gas-phase metallicity, even those where we only could measure an upper limit. The median value for the 30 galaxies where the S/N in [NII] was high enough is 8.35, very close to the median metallicity of the young population stars, as derived from the SED fitting (8.41).

\section{Results and discussion}
\label{sedfit_results}
  \begin{figure*}

   \centering
   \subfigure{\includegraphics[width=0.45\textwidth,keepaspectratio]{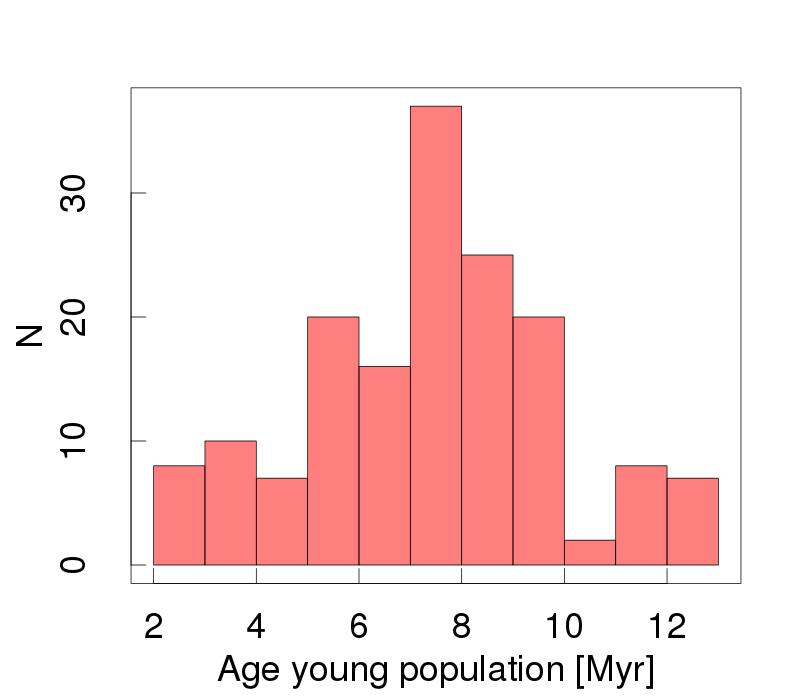}}
   \subfigure{\includegraphics[width=0.45\textwidth,keepaspectratio]{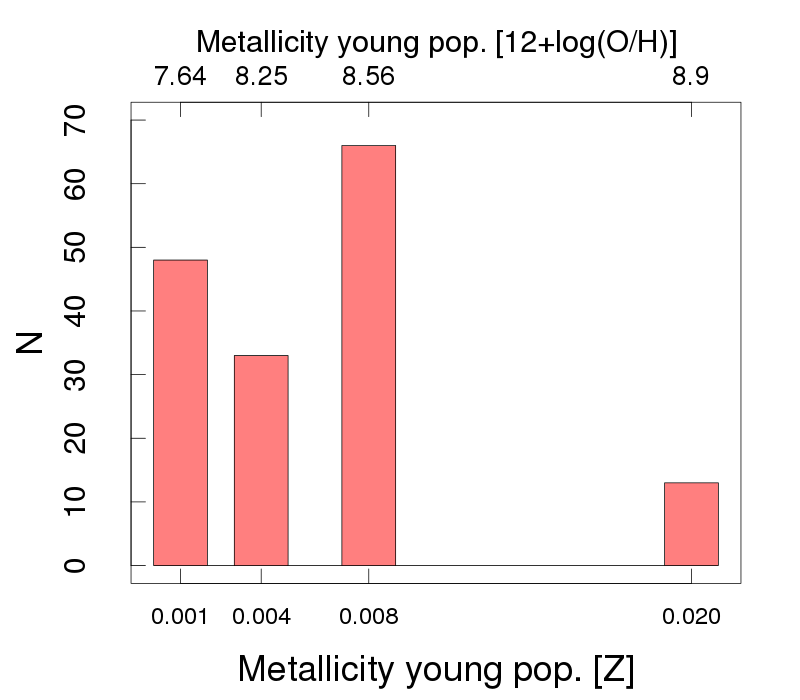}}
   \subfigure{\includegraphics[width=0.45\textwidth,keepaspectratio]{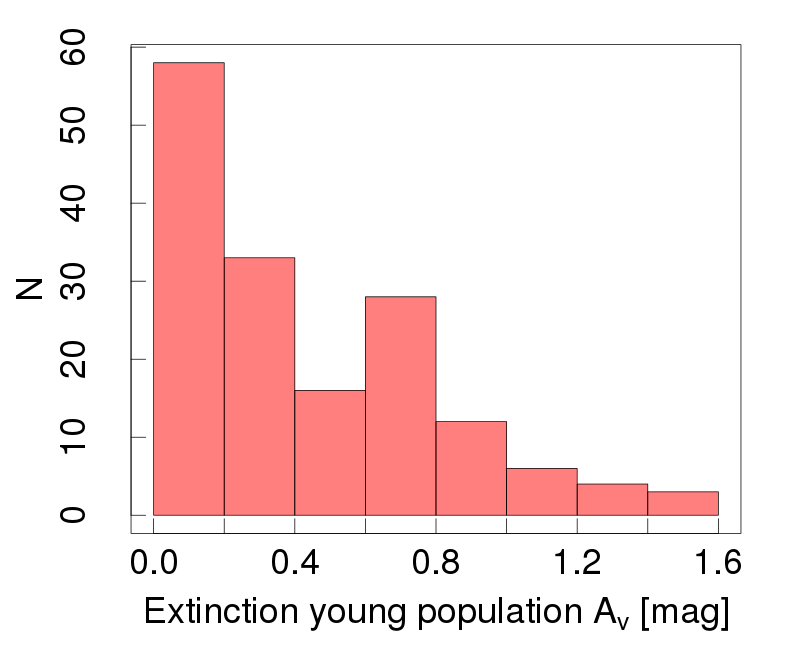}}
   \subfigure{\includegraphics[width=0.45\textwidth,keepaspectratio]{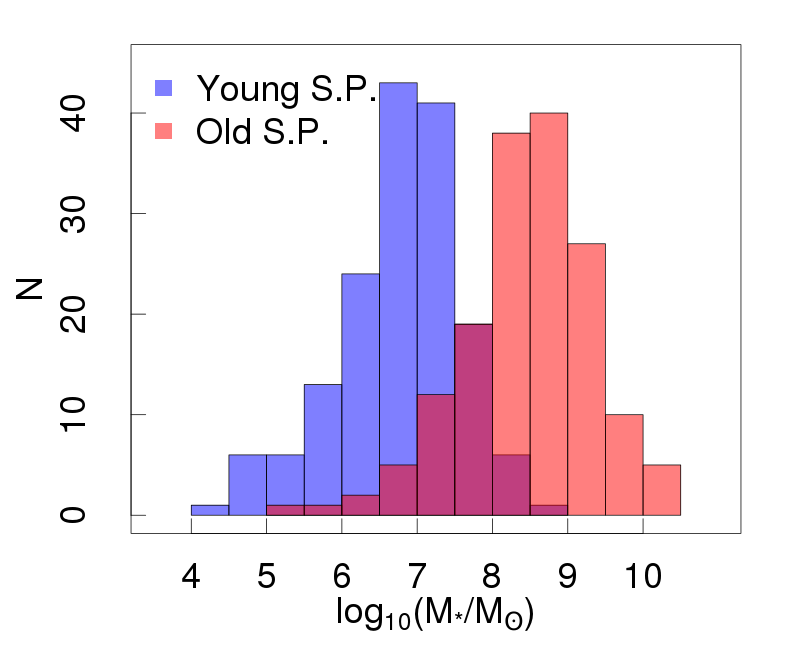}}
      \caption{Stellar population parameters of the young population for the ELG sample, obtained from the SED fitting. \textit{Top left}: Distribution of age of the young stellar population, in millions of years. \textit{Top right}: Distribution of metallicity of the young stellar population, both as a fraction of heavy elements Z and as oxygen abundance 12+log(O/H). \textit{Bottom left}. Distribution of extinction of the young stellar population, in $A_{V}$ magnitudes. \textit{Bottom right}: Distribution of stellar mass of both young and old stellar populations (S.P.), in blue and red, respectively.}
         \label{fig:hist_mass_you}

   \end{figure*}

In what follows we summarize the results of the two population SED fitting on the 160 ELG sample selected by [OIII]5007 and H$\alpha$ detection. We will also review the main gas phase properties of the sample, comparing them and placing them in the context of local and higher redshift surveys. We further provide the color of the galaxies in the sample, as well as their sizes and densities of massive stars. Photometry-derived properties (EW, fluxes, $H\alpha$ luminosity, etc.) are compiled in Table \ref{tab:main_table} for all galaxies in the sample. SED-fitting-derived properties are presented in Table \ref{tab:sed_table}.

\begin{table*}[t]
\renewcommand{\arraystretch}{1.5}
  \centering
\caption{Results from the SED fitting. Fixed values for the old population: Age (2 Gyr), metallicity (Z=0.004) and extinction (E(B-V)=0.08). The complete table is available online; only the first rows are shown here as guidance.}
\begin{tabular}{lllllllllll}
\hline\hline
ID & $ EW_{H\alpha}$& $\chi^{2} $& $ Age_{y}$& $Z_{y}$&$ A_{V.y}$& $M_{y}$& $ M_{o}$ & $log(N_{O\star})$ & R & $\sigma_{O\star} $\\ 
(1) & (2) & (3) & (4) & (5) & (6) & (7) & (8) & (9) & (10) & (11) \\
\hline

   20002280 & 61 & 1.09 & $10.0\substack{+0.35 \\ -2.1}$& $0.001\substack{+0.005 \\ -0.0}$ & $0.37\substack{+0.17 \\ -0.087}$ & $7.1\substack{+0.1 \\ -0.2}$ & $8.9\substack{+0.0 \\ -0.2}$ & 3.9 & 4.88 & 99 \\ 
  10000098 & 23 & 0.85 & $2.5\substack{+5.75 \\ -0.0}$ & $0.004\substack{+0.012 \\ -0.0}$ & $0.00\substack{+1.0 \\ -0.0}$ & $5.0\substack{+2.0 \\ -0.0}$ & $9.2\substack{+0.0 \\ -0.2}$ & 2.4 & 2.03 & 21 \\ 
  10000145 & 42 & 0.61 & $7.5\substack{+3.3 \\ -0.55}$ & $0.008\substack{+0.0 \\ -0.006}$ & $0.50\substack{+0.34 \\ -0.26}$& $6.6\substack{+0.6 \\ -0.2}$ & $8.5\substack{+0.1 \\ -0.2}$ & 3.5 & 3.08 & 100 \\ 
  10000515 & 38 & 2.49 & $12.5\substack{+0.0 \\ -0.0}$ & $0.001\substack{+0.0 \\ -0.0}$ & $0.87\substack{+0.01 \\ -0.0}$ & $7.7\substack{+0.01 \\ -0.01}$ & $8.7\substack{+0.04 \\ -0.01}$ & 3.7 & 5.52 & 49 \\ 
  10000777 & 34 & 1.09 & $6.5\substack{+0.7 \\ -0.7}$ & $0.004\substack{+0.003 \\ -0.0}$ & $0.12\substack{+0.087 \\ -0.0}$ & $7.3\substack{+0.2 \\ -0.08}$ & $10.0\substack{+0.03 \\ -0.03}$ & 4.5 & 3.95 & 609 \\ 
   & & & & ......& & & & & & \\
\hline
\multicolumn{8}{l}{(1) SHARDS ID.} \\
\multicolumn{8}{l}{(2) Photometrically derived H$\alpha$ EW, in \text{\AA}ngstroms.}\\
\multicolumn{8}{l}{(3) Reduced chi-squared of the best-fit.}\\
\multicolumn{8}{l}{(4) Age of the young stellar population, in Myr.}\\
\multicolumn{8}{l}{(5) Metallicity of the young stellar population.}\\
\multicolumn{8}{l}{(6) Extinction ($A_V$) of the young stellar population, in magnitudes. }\\
\multicolumn{8}{l}{(7) Mass of the young stellar population, in log(M$_{\odot}$.)}\\
\multicolumn{8}{l}{(8) Mass of the old stellar population, in log(M$_{\odot}$.)}\\
\multicolumn{8}{l}{(9) Number of O stars per galaxy (logarithm). }\\
\multicolumn{8}{l}{(10) Physical effective radius of the galaxy in kpc, based on F160W data \citep{2014ApJS..214...24S}.}\\
\multicolumn{8}{l}{(11) Surface density of O stars, in N$_{\star}$ \ kpc$^{-2}$.}\\
 \end{tabular}

\label{tab:sed_table}

\end{table*}

\subsection{The host and the burst: results from the fit of the SED}

Figure \ref{fig:hist_mass_you} shows the distribution of the two stellar populations' parameters that best reproduce the SED of our ELG sample. A summary of the best-fitting values for each galaxy and their uncertainties are given in Table \ref{tab:sed_table}. As shown in the examples of Fig. \ref{fig:ejemplos_SEDS}, several different cases exist, with a range of H$\alpha$ and [OIII]5007 EW, extinctions, and relative strengths of the young and old populations.

For the vast majority of the galaxies, we are able to successfully fit their SED with a simple two-populations model, using effectively only four parameters (age, extintion, metallicity and mass of the young population). This means that we can fit these galaxies with a physically motivated model (many studies have found an underlying old host galaxy and young star-forming regions), but reducing the number of free parameters with respect to the usual exponential SFH model. We consider this approach more robust given both the nature of the data and the strong covariances within the parameters involved in the fit.

Regarding the ages of the young stellar populations in
panel (a) of Fig. \ref{fig:hist_mass_you}, we obtain a median value of 8 Myr, and an extended distribution ranging from 2.5 to 13 Myr.

The metallicities of the young stellar population (Fig. \ref{fig:hist_mass_you}, panel b) are low, $\sim 50\%$  of them being $Z=0.004$ (12+$\log$(O/H)=8.25) or lower, and $\sim 92\%$ of them being $Z=0.008$ (12+$\log$(O/H)=8.56) or lower. We compute the corresponding stellar oxygen abundances directly from the models \citep{1992A&AS...96..269S,1993A&AS..101..415C,1993A&AS...98..523S}. Comparing these values to the compilations presented in Fig. 11 of \cite{2016ApJ...830...64B} and Fig. 8 of \cite{2017MNRAS.467.3759T}, we find that the SF bursts of the galaxies in our sample have typical metallicities ranging from that of Sextans A to a value between the Small and Large Magellanic Clouds.

The extinction of the young population (panel c) tends also to be small, with median $A_V \sim 0.37$ mag. These values are low compared to some other surveys where extinction was computed for each source \citep{2008ApJ...677..169V,2014ApJ...784..152A}. It is also low compared to the extinction of 1 magnitude typically applied in other surveys \citep[e.g., in ][]{2013MNRAS.428.1128S}. This small extinction is easily understandable given the low masses of the galaxies in our sample \citep{2013ApJ...763...92Z}. For a sample with similar values of stellar mass to those in ours, \cite{2016A&A...591A.151G} find similar values for the extinction although at higher redshift. The low metallicity and small extinction are consistent together, and are also consistent with the mass values, given the mass-metallicity relation \citep{2012ApJ...757...54Z,2013ApJ...763...92Z,2010MNRAS.409..421G}.

The mass of both the young and the old stellar populations (Fig. \ref{fig:hist_mass_you} panel d) span a wide range of values: From $10^5 M_{\odot}$ to 
$10^8 M_{\odot}$ being the young one and from $10^7 M_{\odot}$ to $10^{10} M_{\odot}$ being the old one in most cases. 

The differences with higher redshift studies are partially due to our improved sensitivity to lower-mass galaxies. For example, when considering only galaxies with $M_{\star}/M_{\odot} > 10^{8.43} M_\odot$ \citep[around the detection limit in][]{2015ApJ...812..155C}, our median mass grows to $10^{8.96} M_\odot$, closer to their median value of $10^{9.53} M_{\odot}$ (at higher redshift, $z=0.84$). In addition, the relative lack of high-mass galaxies in our sample is due to the small volume covered by the survey at low redshift, and also to the decreasing fraction of high-mass star-forming galaxies at lower redshift \citep{2005ApJ...619L.135J,2009ApJ...690.1074M}. When comparing masses of other studies to ours, we have taken into account the difference between their \cite{2003PASP..115..763C} and our \cite{1955ApJ...121..161S} IMF mass normalization, multiplying by 1.7 (or adding 0.23 dex in logartihmic scale) the \cite{2003PASP..115..763C} mass values.

\begin{figure}
  \centering
    \includegraphics[width=0.45\textwidth,keepaspectratio]{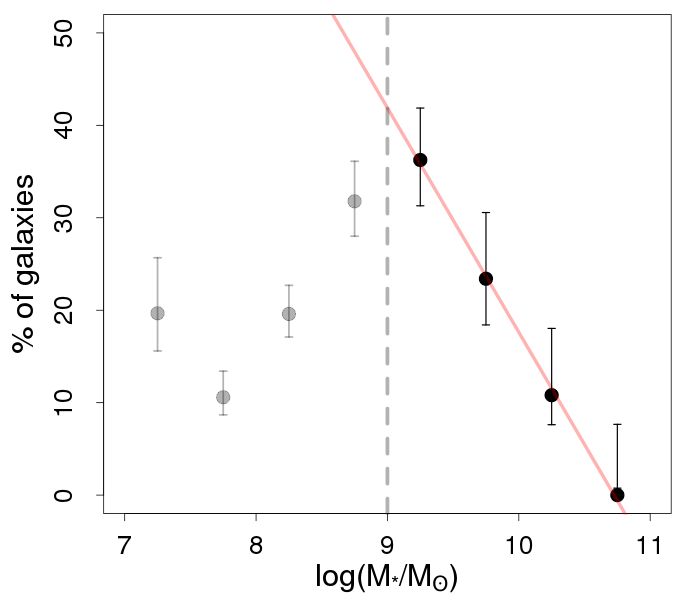}
    \caption{Percentage of ELG compared to the reference sample, as a function of stellar mass. We reach our completeness limit at 10$^9$ M$_{\odot}$ (vertical dashed gray line). The red line represents the best fit to the significative data points. Data points under the completeness limit are shown in gray. }
         \label{fig:percent_mass}
   \end{figure}

Figure \ref{fig:percent_mass} presents the relation between the percentage of ELG compared to the reference sample as a function of the total stellar mass of the galaxies. This fraction grows with decreasing stellar mass, until we reach the completeness limit ($\sim 10^9 M_{\odot}$). This behavior is consistent with previous results (e.g., \citealt{1997ApJ...481...49H}, \citealt{2016A&A...590A..18R} and \citealt{2011MNRAS.411..675S}). Comparing with Figure 4 in \cite{2011MNRAS.411..675S}, we find that the slope of the relation is compatible, despite the different mass range and redshift.

Finally, the galaxies with spectra are more massive (median mass of 10$^{8.81}\ M_{\odot}$) than galaxies without spectra (10$^{8.05}\ M_{\odot}$). This is due to the improved depth of SHARDS data compared to spectroscopic surveys.

\subsection{Gas phase properties}
\label{gasphase}
  \begin{figure}
  
   \centering
   \subfigure{\includegraphics[width=0.47\textwidth,keepaspectratio]{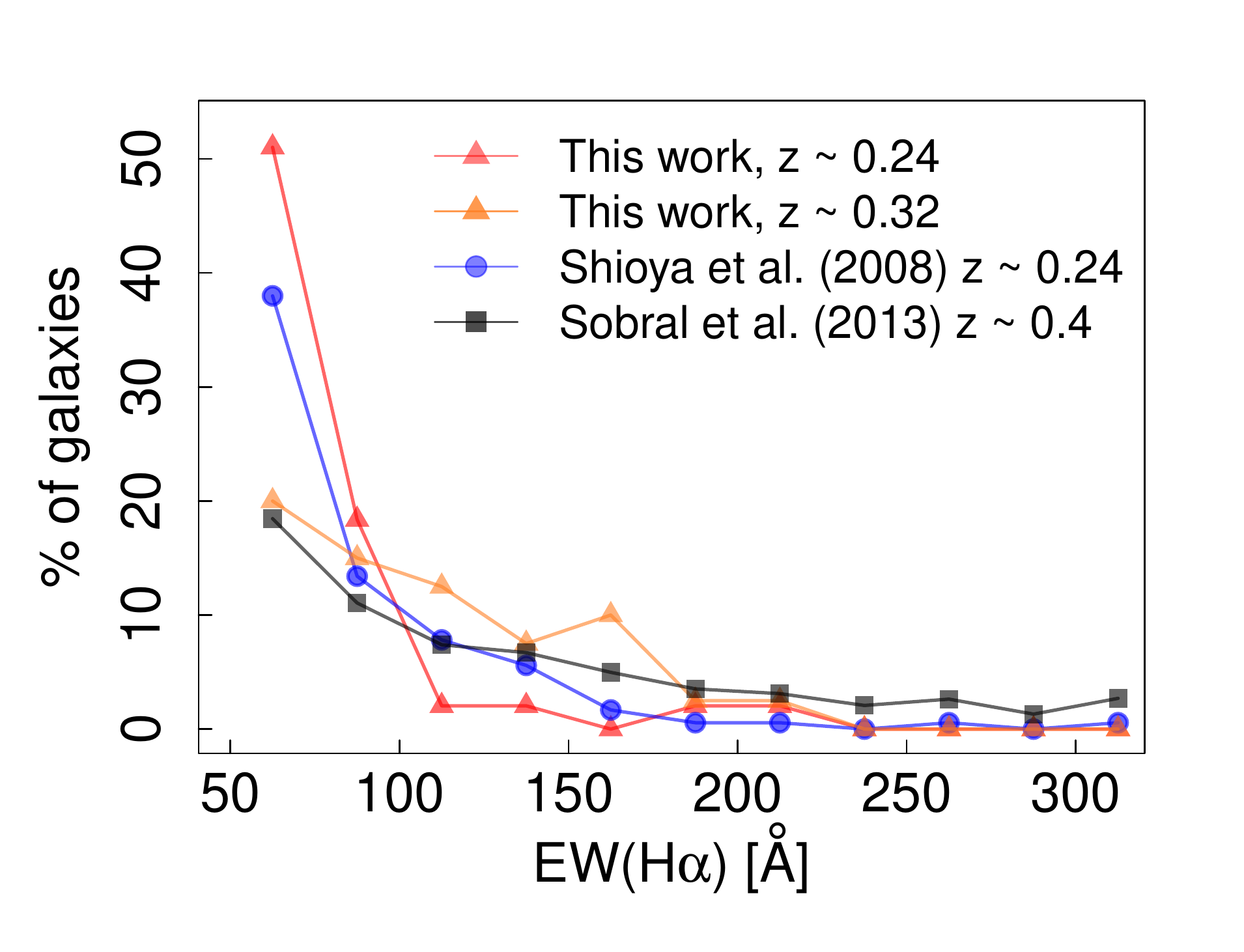}}
   \subfigure{\includegraphics[width=0.47\textwidth,keepaspectratio]{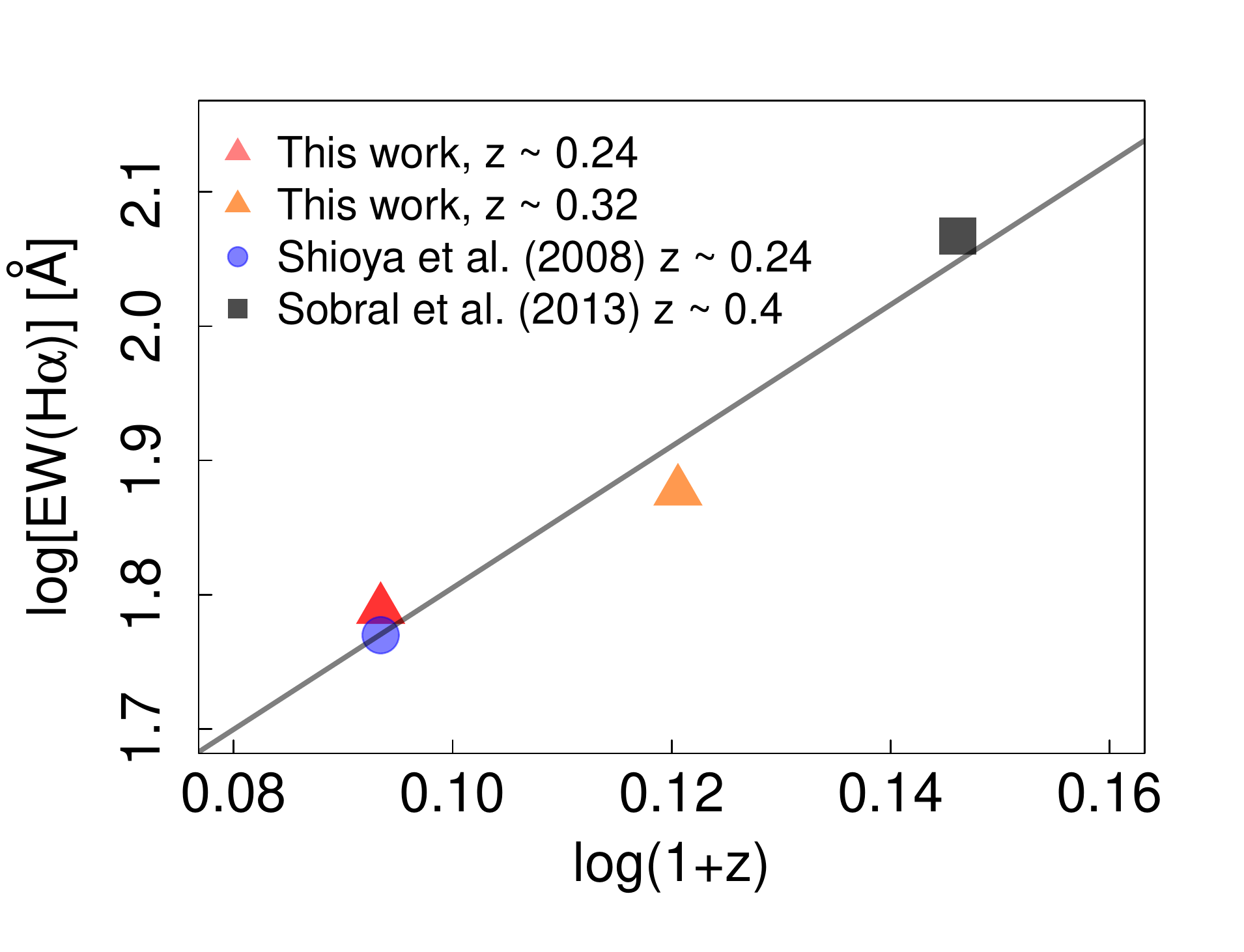}}
      \caption{\textit{Top:} Percentage of galaxies in a given H$\alpha$ EW bin for different samples. As redshift increases, distributions tend to have larger EW. \textit{Bottom:} Median value for the H$\alpha$ EW in different samples as a function of redshift. The gray line represents the best fit. For our sample (triangles), we consider two redshift bins: $z\sim$ 0.24 (red) and $z\sim$ 0.32 (orange). We also plot \cite{2008ApJS..175..128S} as blue circles and \cite{2013MNRAS.428.1128S} as gray squares. To ensure that all samples are complete and comparable, we consider only  galaxies with H$\alpha$ EW>50 \text{\AA}.}
         \label{fig:compar1}
         
   \end{figure}  

The rest-frame EW of H$\alpha$ and [OIII]5007 were presented in Fig. \ref{fig:EW}, showing median values of 56 \text{\AA} and 35 \text{\AA}, and with 50\% of the samples with 40 \text{\AA} $\leq$ EW$_{H\alpha} \leq$ 82 \text{\AA} and 22 \text{\AA} $\leq$ EW$_{[OIII]} \leq$ 60 \text{\AA}, respectively. We compared in Fig. \ref{fig:EWo3} the values derived from SHARDS SEDs and from TKRS and DEEP2 spectra in the subsample of galaxies where they were available. We find a good match in photometrically and spectroscopically derived EW for both lines, with 96\% of the deviations being smaller than 3$\sigma$. Some points have a large difference, and this is probably due to differences in aperture, where the slit did not cover the whole galaxy (or the whole star-forming region).

In Fig. \ref{fig:distredshift} we present the amount of ELGs per redshift bin as a percentage of the reference sample. Our result is consistent with that reported by \cite{1997ApJ...481...49H} for galaxies in our redshift bin, $\sim$ 25\%. Their work was based on a low-resolution spectroscopic survey, dealing with more massive galaxies, but with a lower EW detection limit than us (15 {\AA} in the OII[3727] line). More recently,  \cite{2016A&A...590A..18R} found a higher fraction ($\sim$ 45\%), dealing also with massive galaxies ($\sim$ 10$^{10}$ M$_{\odot}$) and with a much lower EW limit (10 {\AA} in the H$\alpha$ line).

We compare with other surveys at similar redshifts with available EWs, \cite{2008ApJS..175..128S} and \cite{2013MNRAS.428.1128S}. We use only galaxies with EW > 50 \text{\AA} to ensure the sample is complete and comparable. Figure \ref{fig:compar1} shows the distribution of EW (top panel) and its median value as a function of redshift (bottom panel). We split our sample into two redshift bins, and show that our results agree well with the redshift evolution of EW. \cite{2015ApJ...812..155C} present the redshift evolution of the median [OII]3727 EW (their Figure 11), which also follows a linear relation, but with a lower slope ($\alpha$=3.2) than ours ($\alpha \sim$ 5).

\begin{figure}
   \centering
   \includegraphics[width=0.5\textwidth,keepaspectratio]{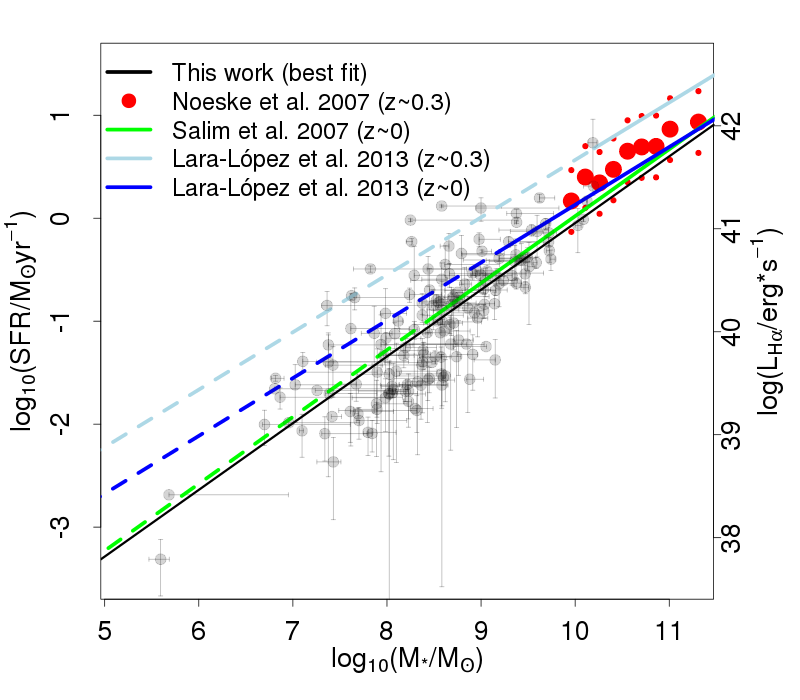}
      \caption{SFR of the ELG sample vs. stellar mass, in gray dots. The black line represents the best fit to our data. In light and dark blue, the relations presented in \cite{2013MNRAS.434..451L} for z$\sim$ 0.3 and z$\sim$ 0 galaxies, respectively. The green line corresponds to the linear fit to \cite{2007ApJS..173..267S} data. The former three lines are continuous in the mass range where the respective samples are complete, and dashed otherwise. In large red dots, the median values for the z$\sim$0.3 \cite{2007ApJ...660L..43N} sample (in smaller dots, the 1$\sigma$ contours).}
         \label{fig:ssfr_mass}
   \end{figure}

Using the H$\alpha$ flux computed in Section \ref{detection_lines} we derive the H$\alpha$ luminosity of the galaxies, and use it to estimate the SFR using \cite{1998ARA&A..36..189K} calibration. We correct both quantities for dust extinction using the values derived in the SED fit for the young stellar population. We note that we use the stellar extinction in this procedure as a first order aproximation to the actual nebular extinction, which may be different \cite[e.g.,][]{2016ApJ...828..108R}. We cannot directly estimate the nebular extinction spectroscopically, since not all our galaxies have spectral information, and those where it is available are not flux callibrated, preventing us from using the Balmer decrement method. In Fig. \ref{fig:ssfr_mass} we show the SFR-mass relation for our sample (also known as "star formation main sequence").
We have extended the usual SFR-mass relation to the region of both low mass and low SF in galaxies at low redshift, found to be only sparsely populated in previous analyses. Those studies based on large samples \citep{2007ApJS..173..267S,2007ApJ...660L..43N,2013MNRAS.434..451L} only presented information for a few galaxies under 10$^9$ solar masses. Those focused on low-mass galaxies \citep{2014ApJ...780..122L,2015A&A...578A.105A,2017A&A...601A..95C} are biased towards highly star-forming systems. Our data is compatible with the extrapolation to lower masses of some local samples \citep{2007ApJS..173..267S,2013MNRAS.434..451L}, and lies slightly under the values computed by the GAMA team for their subsample at $0.23 < z < 0.36$ \citep{2013MNRAS.434..451L}. This GAMA sample, however, lies slightly above the \cite{2007ApJ...660L..43N} data, within the 1$\sigma$ scatter. We find that the slope in the relation is consistent with that derived in previous, higher-mass analyses, considering the uncertainties. We do not see the strong flattening of the relation at lower masses claimed by \cite{2013ApJ...772...48P}. This suggests that, overall, the star-forming mechanisms operating in high-mass galaxies are also in play in their lower-mass counterparts. We have normalized the \cite{2003PASP..115..763C} IMF mass values of these studies to \cite{1955ApJ...121..161S} values.

\begin{figure}
  \centering

   \includegraphics[width=0.45\textwidth,keepaspectratio]{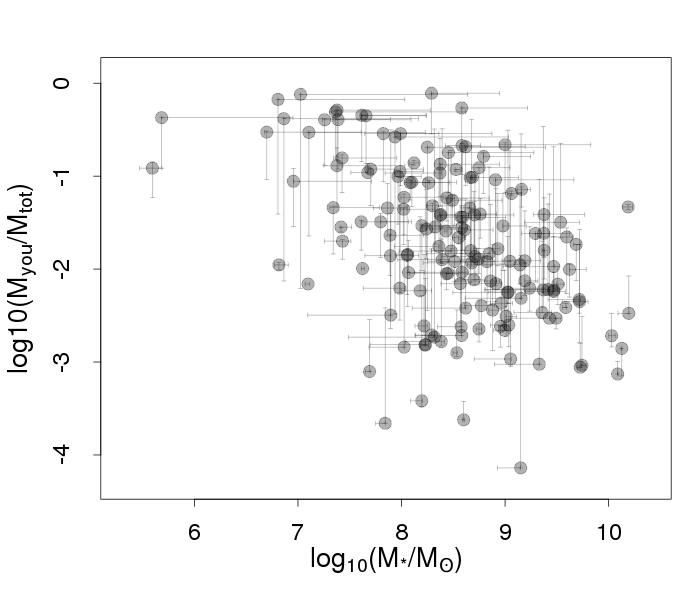}
\caption{Burst strength (mass of the young population divided by total stellar mass) as a function of total stellar mass in the galaxies of our sample. The median Spearman correlation factor, taking into account the uncertainties, is -0.47. }
 \label{fig:burst_vs_mass}
    \end{figure}

We found that the scatter of the SFR-mass relation (Fig. \ref{fig:ssfr_mass}) strongly depends on the burst ratio (mass of the young stellar population divided by the total mass of the galaxy), with higher-burst-ratio galaxies located in the upper region of the relation. Figure \ref{fig:burst_vs_mass} shows the burst ratio as a function of the total stellar mass, and we found that lower-mass galaxies present higher burst ratios than higher-mass galaxies. We ran a series of 10000 bootstrap and Monte Carlo simulations and in all of them the Spearman correlation factor was negative ($p<10^{-4}$), with a median value of -0.47. This correlation is consistent with previous studies finding low-mass galaxies to have more burst-like SF; see for example \cite{2010MNRAS.405.2594G},  \cite{2013MNRAS.434..209B}, and \cite{2013ApJ...770...57B} for observational evidence, and \cite{2015MNRAS.450.4486F} and \cite{2016ApJ...833...37G} for results based on simulations. This result independently supports our choice of a SED model with a recent star-forming burst rather than an exponentially declining SFH. 

In order to check the consistency of our SFR derivation, we compare our values to those obtained using infrared data gathered from the Rainbow multiwavelength database\footnote{\url{http://rainbowx.fis.ucm.es/Rainbow_navigator_public/}} \citep{2011ApJS..193...13B,2011ApJS..193...30B}. We use the total infrared luminosity, $L(TIR)$ (from 8 to 1000 $\mu$m), computed from synthetic spectra using Rainbow SED fits. These fits take into account observational data from the UV (GALEX) to the far-infrared, gathering data from both $\it{Spizter}$ and $\it{Herschel}$ . We then apply \cite{2015A&A...584A..87C} recipes for computing the SFR based on different tracers. We consider, in particular, 
\begin{equation}
SFR(M_{\odot}\ yr^{-1}) = 5.5 \times 10^{-42} \ [L(H \alpha _{obs}) + 0.0024 \times L(TIR)]
,\end{equation}
\begin{equation}
SFR(M_{\odot}\ yr^{-1}) = 5.5 \times 10^{-42} \times L(H \alpha _{corr})
,\end{equation}

where $L(H \alpha _{obs})$ and $L(H \alpha _{corr})$ are the measured and extinction-corrected $H\alpha$ luminosities, respectively, computed from our photometric $H\alpha$ flux measurements. The two resulting SFR estimations agree within 3$\sigma$ for 85\% of the sample. This is a remarkable agreement, given the multiple sources of uncertainty and assumptions made. This result also gives an independent confirmation of both our $H\alpha$ flux measurement and the young population extinction derived from the SED fitting, described in section \ref{SEDfit}.

We must note that the assumption of an instantaneous star-forming burst and the derivation of a SFR are, sensu stricto, inconsistent (see \citealt{2010A&A...511A..61O}). Nevertheless, as most other works in literature do, it is interesting to derive SFR estimations both for comparison purposes, and as an independent estimation of the mean SF in the galaxies during the most recent $\sim$ 10 Myr.

In Sect. \ref{spectra} we also used the spectra to compute the gas-phase metallicity using \cite{2013A&A...559A.114M} calibration for the [NII]/H$\alpha$ ratio in the 26 galaxies that have S/N>3 in [NII] (upper limits were determined for the rest of the spectroscopic sample). The median value is $\log$(O/H)=8.36 as seen in Fig. \ref{fig:met_gas_cumul}, and the median mass of this sub-sample is $10^{9.31} M_{\odot}$. All gas phase measurements show sub-solar metallicities, as well as all 92\% of the SED derived young stellar populations. Given the sparse sampling of metallicity given by the S99 models, we cannot further compare the stellar metallicity and gas metallicity values. Taking into account the systematic shift between different metallicity estimators studied in \cite{2008ApJ...681.1183K}, the gas metallicity of our sub-sample agrees with the expected metallicity at the observed mass and redshift range \citep{2013ApJ...763....9Y, 2016MNRAS.463.2002H}.

\subsection{The color of the galaxies}
  \begin{figure}
   \centering
   \includegraphics[width=0.5\textwidth,keepaspectratio]{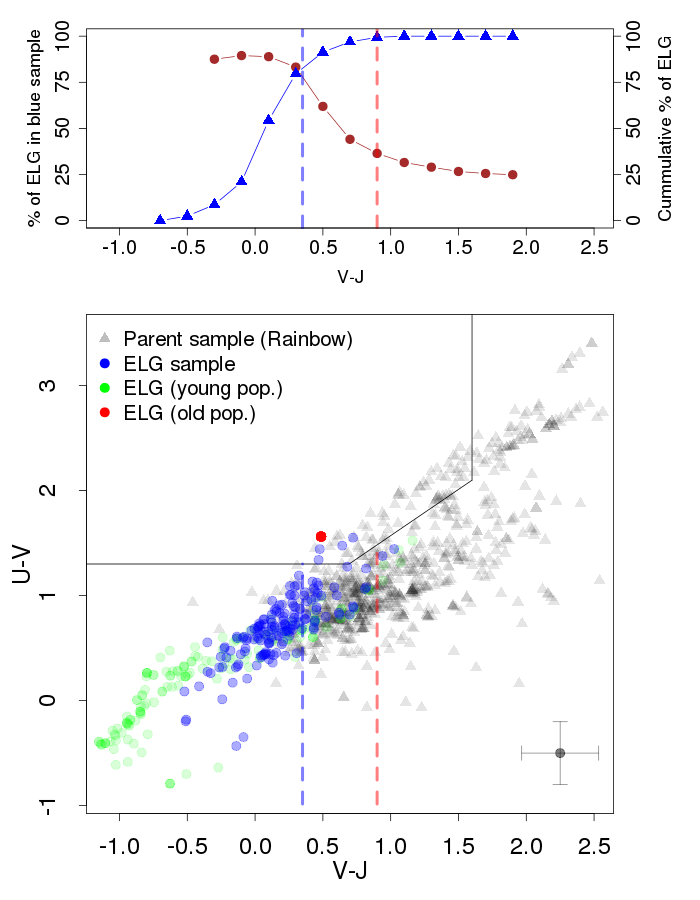}
      \caption{\textit{Top}: In blue, cumulative percentage of ELG galaxies as a function of V-J rest-frame color. In brown, cumulative percentage of ELG in the blue sample as a function of V-J color. Vertical dashed lines are the proposed V-J values to select ELG in photometric surveys (see text). \textit{Bottom}: UVJ diagram showing the galaxies in our sample along with a representative reference sample from the Rainbow database (gray triangles). We represent the color of the galaxies in our sample in blue dots, from our SED fitting. We plot also the color of the old and young populations in red and dark green dots, respectively. Limits for the quiescent region (top-left region) are taken from \cite{2011ApJ...735...86W} for $z$<0.5. Vertical dashed lines as in top panel.}
         \label{fig:colors}
   \end{figure}

Color-color diagrams have been historically used to separate star-forming and quiescent galaxies \citep[see e.g.,][]{1996MNRAS.283.1388M}. More recently, the UVJ diagram \citep{2005ApJ...624L..81L,2007ApJ...655...51W} has proven to be useful for that purpose at different redshifts. In this section we explore the position of our sample of galaxies in the UVJ plane, to test whether they would have been identified in a color selection, and if we could define a more restrictive criteria to select emission line galaxies. Using the fluxes derived from the SED fitting, we computed the synthetic UVJ magnitudes of the ELGs, and also those of the two modeled populations (see bottom panel of Fig. \ref{fig:colors}). As expected, almost all galaxies in our sample lie in the blue region of the diagram \cite[according to the ímits defined in][]{2011ApJ...735...86W}, with the young populations in bluer regions of that area. Old populations lie in the quiescent area (since we used a fixed age, metallicity, and extinction for the old stellar population, all have the same color).

To compare with the non-emission-line galaxies, we also plot the reference sample from
SHARDS in the bottom panel of Fig. \ref{fig:colors}; the synthetic colors are taken from the Rainbow database \citep{2011ApJS..193...13B,2011ApJS..193...30B}. Given the different methods of SED fitting, we notice a scatter and a small systematic error between the value of the colors using our SED fitting method and that of Rainbow for the ELG sample. We represent the typical scatter as error bars in the bottom right corner of the bottom panel of figure \ref{fig:colors} (the systematic error is smaller than the scatter). Using the Rainbow SED fitting for our ELG sample would not change the conclusions of this section. 

All but four of the galaxies in the ELG sample lie in the blue region of the diagram. It is noticeable that our ELG sample occupies a different region in the UVJ diagram compared to the whole sample of blue galaxies.
To check the statistical significance of this difference, we use the Anderson-Darling criterion \citep{CIS-74828} for both U-V and V-J colors. In both cases, the p-values are much smaller than 0.05, so we can confidently claim that they are two different populations. 
 
Given this result, we find new color cuts to select ELGs in broad band surveys. Selecting blue galaxies with V-J < 0.35, 84\% of them are ELG, and they amount to 76\% of the total ELG sample. On the other hand, limiting to V-J > 0.9 results in selecting a blue galaxy sample with no emission (only 1\% ELG according to our results) while keeping 52\% of the non-ELG sample. In between those color cuts, both populations are mixed. The cumulative percentage of ELG galaxies as a function of V-J color is presented in blue triangles in the top panel of Fig. \ref{fig:colors}. The cumulative fraction of galaxies in the blue region of the diagram that are ELG is plotted in brown dots; it reaches a plateau at 25\% for high values, since that is the percentage of blue galaxies that are ELG.

More generally, the presence of a large number of blue galaxies for which we do not detect emission lines is understandable, given the observational limits on EW detection, and the persistance of galaxies in the blue region of the diagram after a SF event. Considering the mass ratio of young to old populations of our sample, $\sim$ 50\% of the galaxies would still be in the blue region of the diagram 100 Myr after the burst.

\subsection{Density of O-type stars and feedback}

We use the higher spatial resolution 3D-HST data (the \textit{flux radius} parameter from its photometric catalog) to estimate the physical sizes of the galaxies. All galaxies are well resolved in HST images, given that the lower detectable radius would be $\sim$ 0.3 kpc. No statistical difference is detected between the ELG and the reference sample.

The amount and density of massive stars (O type, fundamentally) in a galaxy is related to the mechanical energy that the SF burst is depositing into the interstellar medium. Very high densities can create galactic superwinds \citep{1998MNRAS.293..299T,2010ApJ...708.1621T} that inhibit further SF and contaminate the intergalactic medium (IGM) with enriched material. Other models however predict the possibility that young massive clusters face an intense cooling by frequent interactions between the nearby winds of individual stars. This would result in a solution of positive feedback if much of the - otherwise ejected - material were to be cooled down and kept within the cluster volume to produce more stars \citep{2005ApJ...620..217T}.

\begin{figure}
   \centering
  \includegraphics[width=0.47\textwidth,keepaspectratio]{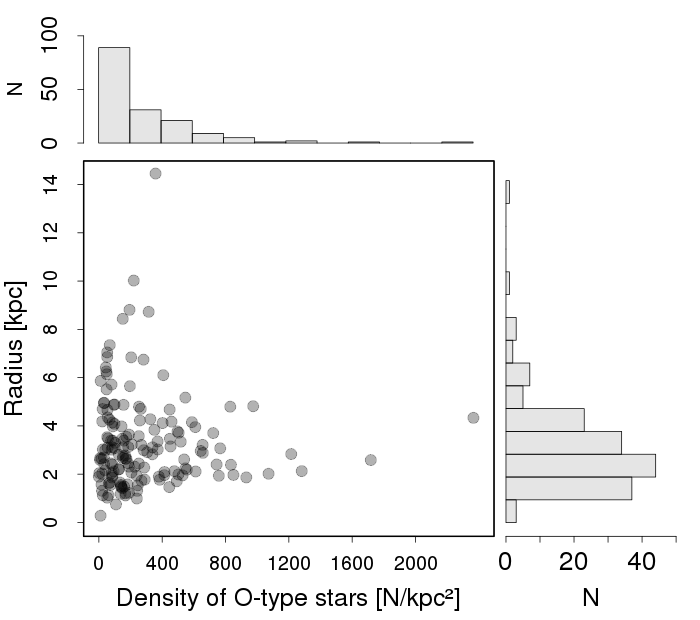}
   \caption{Radius and density of O-type stars of the ELG sample. \textit{Main panel:} Radius of the galaxies as a function of O-type stars surface density. \textit{Top:} Histogram of O-type stars surface density in the ELG sample, computed using the SED fitting results \textit{Right:} Histogram of galaxy effective radius, computed from \textit{HST} infrared data (F160W filter).}
   \label{fig:ostars_rad}
 \end{figure}

Using the mass, metallicity, and age of the young stellar population derived from the SED fit, we retrieve  the amount of O stars per galaxy from the Starburst99 output. Then, using the radius of the galaxy taken from the 3D-HST catalog we can estimate the surface density of O stars in each galaxy. In Fig. \ref{fig:ostars_rad} we present the histograms and scatter plot of those variables. Most galaxies show radii smaller than 6 kpc and densities lower than 400 stars per kpc². The largest galaxies present low O-star densities, which is likely due to the fact that SF tends to happen in localized regions, not over the whole galaxy, and therefore by using the global galaxy size we are only providing an upper limit. There is also a small population of galaxies with very high O-star densities that present small radii. They have over 1000 O-stars per kpc², reaching the order of magnitude of individual starbursts, such as the 30 Doradus HII region (also known as Tarantula Nebula) in the Large Magellanic Cloud, that presents approximately 5000 O-stars per kpc² in the central 300 pc \citep{2013A&A...558A.134D}. The simulations presented in \cite{2005ApJ...620..217T} show that the shape and components of the H$\alpha$ emission line in young, massive clusters could be used to determine the feedback regime (positive, negative, or bimodal) that the cluster is experiencing. The densest galaxies in our sample could therefore be potential targets for high-resolution spectroscopy.

\section{Conclusions}
\label{conclusions}

We identify 160 emission-line galaxies in the SHARDS survey on the GOODS-N field, up to $z$=0.36, via simultaneous detection of the H$\alpha$ and [OIII]5007 emission lines, reaching low-stellar-mass and low-SFR galaxies thanks to SHARDS depth and spectral resolution. We developed a new algorithm optimized to find emission lines in SHARDS multi-filter medium-band photometry. By selecting galaxies with both lines we avoid contamination caused by interlopers at other redshifts. The continuous wavelength coverage of the survey allows for a precise determination of the continuum, resulting in robust measurements of the EW and flux of the emission lines.

We detect faint emission lines, reaching limits of $\sim$ 15 \text{\AA} in EW and $ \sim 4 \times 10^{-18} $ erg\ s$^{-1}$\ cm$^{-2}$ of flux in H$\alpha$, with 50\% completeness at 35 \text{\AA} and $7.4 \times 10^{-17} $ erg\ s$^{-1}$\ cm$^{-2}$. These values are similar to the limits found in \cite{2015ApJ...812..155C}, and comparable to those of narrow-band surveys \citep{2013MNRAS.428.1128S} with wider redshift coverage. We identify as ELG $\sim$ 20\% of the galaxies in the reference sample (all galaxies detected in SHARDS with the same redshift and absolute magnitude limits), compatible with previous results. This percentage decreases with mass at a similar rate as seen in \cite{2011MNRAS.411..675S}. For high EW galaxies, the percentage we detect is approximately two times greater than what \cite{2016A&A...592A.122H} detect in COSMOS, probably caused by SHARDS depth and continuous redshift coverage.

Using the photometrically derived EW in H$\alpha$ as a constraint, and using ancillary data from ALHAMBRA and GALEX, we successfully fit the SED of the galaxies in the sample using a model with two single stellar populations. With fixed age, extinction, and metallicity in the old stellar population, we find robust results for the young stellar population properties, even considering different ages for the old population. 

The age of the young stellar population is low, with a median value of 8 Myr, and shows an extended distribution from 2.5 to 13 Myr. Masses of the young populations range from $10^6 M_{\odot}$ to $10^8 M_{\odot}$ (with a median of $10^{6.9} M_{\odot}$) and those from the old one range from $10^8 M_{\odot}$ to $10^{10} M_{\odot}$ (with a median of $10^{8.5} M_{\odot}$). As a result, the burst strength goes from 0.001 to 0.1. The metallicity of the young stellar population is low, with $\sim 92\%$ of the sample presenting 12+$\log$(O/H) $\leq$ 8.56. The extinction of the young populations is also small, with a median value of $A_V \sim 0.37$ mag. Low metallicities and extinctions are known to be correlated together and also with low masses, showing our sample is in agreement with previous studies at low to intermediate redshifts.

The distribution of H$\alpha$ EW is compatible with previously defined trends in our redshift range. Given the abundant spectroscopic coverage in the field, we have analyzed the spectra of the galaxies in our sample where it was available, finding good agreement with photometrically derived EWs. The subsample without spectroscopic coverage is 1.85 magnitudes fainter than the one with spectra, which is one of the advantages of a deep photometric survey such as SHARDS.

The vast majority of gas-phase metallicities derived from the [NII]/H$\alpha$ ratio in the galaxies with spectra are lower than solar metallicity, as well as SED-fitting-derived stellar metallicities. They occupy the expected position in the mass-metallicity relation considering previous literature. 

SFR computed with the H$\alpha$ flux derived photometrically is consistent with the value derived from IR, which shows the robustness of both H$\alpha$ and extinction measurements.

The ELGs show very blue UVJ colors compared with all color-selected galaxies. We suggest a new color cut to select ELG in broadband surveys: V-J<0.35 selects ELG, and V-J>0.9 selects non-ELG. The amount of non-ELG blue galaxies is consistent with our limits in emission-line detection, and with the persistence of galaxies in the blue region of the diagram after SF shuts down ($\sim$ 100 Myr). In addition, the size of the ELGs is similar to that of non-ELGs in the reference sample. These findings suggest that ELGs are a transient phase of the same class of galaxies.

In order to explore the possible feedback regimes in our sample, we find a number of galaxies with a high density of O stars (more than 1000 O stars per kpc²). They could be interesting targets for follow-up high-resolution spectroscopy, to look for the impact of the burst into the IGM.

In forthcoming papers, we will extend this work to higher redshifts and perform a detailed morphological analysis on the host galaxies, as well as on the various properties of the star-forming knots.

\begin{acknowledgements}
This work was partly financed by the Spanish Ministerio de Economía y Competitividad (MINECO) within the ESTALLIDOS project (AYA2013-47742-C4-2P and AYA2016-79724-C4-2-P). AL-C acknowledges financial support from MINECO PhD contract BES-2014-071055 and from grant EEBB-I-16-10913 for a short stay at St. Andrews University. 
J.M.-A. acknowledges support from the European Research Council
Starting Grant (SEDmorph; P.I. V. Wild) and MINECO AYA2013-43-43188-P grant. P.G.P.-G. and A.H-C. acknowledge support from MINECO AYA2015-70815-ERC and AYA2015-63650-P Grants. JMMH acknowledges funding by MINECO grant ESP2015-65712-C5-1-R.
This work is based on observations made with the Gran Telescopio Canarias (GTC), installed in the Spanish Observatorio del Roque de los Muchachos of the Instituto de Astrofísica de Canarias, in the island of La Palma. This work is (partly) based on data obtained with the SHARDS filter set, purchased by Universidad Complutense de Madrid (UCM). SHARDS was funded by the Spanish Government through grant AYA2012-31277. This work is partly based on observations taken by the 3D-HST Treasury Program (GO 12177 and 12328) with the NASA/ESA HST, which is operated by the Association of Universities for Research in Astronomy, Inc., under NASA contract NAS5-26555. This work has made use of the programming software R \citep{rstat}.
We would like to thank S. Barger for kindly providing longslit spectra for 26 galaxies, and to C. Leitherer for help making use of the Starburst99 code.
\end{acknowledgements}


\bibliographystyle{aa}
\bibliography{biblio}

\end{document}